\documentclass[prb,twocolumn,showpacs]{revtex4}
\usepackage{graphicx} 

\usepackage{amsmath,amsthm,amssymb,mathrsfs}
\usepackage{bm}

\begin{document}

\title{Dynamic coupling of ferromagnets via spin Hall magnetoresistance}

\author{Tomohiro Taniguchi}
 \affiliation{
 National Institute of Advanced Industrial Science and Technology (AIST), Spintronics Research Center, Tsukuba, Ibaraki 305-8568, Japan
 }

\date{\today} 
\begin{abstract}
{
The synchronized magnetization dynamics in ferromagnets on a nonmagnetic heavy metal caused by the spin Hall effect is investigated theoretically. 
The direct and inverse spin Hall effects near the ferromagnetic/nonmagnetic interface generate longitudinal and transverse electric currents. 
The phenomenon is known as the spin Hall magnetoresistance effect, 
whose magnitude depends on the magnetization direction in the ferromagnet due to the spin transfer effect. 
When another ferromagnet is placed onto the same nonmagnet, these currents are again converted to the spin current by the spin Hall effect 
and excite the spin torque to this additional ferromagnet, resulting in the excitation of the coupled motions of the magnetizations. 
The in-phase or antiphase synchronization of the magnetization oscillations, 
depending on the value of the Gilbert damping constant and the field-like torque strength, is found in the transverse geometry 
by solving the Landau-Lifshitz-Gilbert equation numerically. 
On the other hand, in addition to these synchronizations, 
the synchronization having a phase difference of a quarter of a period is also found in the longitudinal geometry. 
The analytical theory clarifying the relation among the current, frequency, and phase difference is also developed, 
where it is shown that the phase differences observed in the numerical simulations correspond to 
that giving the fixed points of the energy supplied by the coupling torque. 
}
\end{abstract}

 \pacs{85.75.-d, 75.78.-n, 05.45.Xt, 72.25.-b}
 \maketitle


\section{Introduction}
\label{sec:Introcution}

Dynamic coupling of ferromagnets has been of interest in the field of magnetism. 
The dipole interaction has been the basic interaction to excite the coupled motion of the magnetizations, 
and is applied to magnetic recording \cite{hubert98,kudo10}. 
Another method to realize the coupling is to use the spin-transfer effect \cite{slonczewski96,berger96}, 
where the application of an electric current to ferromagnetic/nonmagnetic multilayers 
results in the magnetization switching and self-oscillation \cite{katine00,kiselev03,grollier03,rippard04,kubota05,tulapurkar05,houssameddine07,taniguchi08,slavin09,kubota13}. 
The coupled dynamics through pure spin current generated in spin pumping \cite{tserkovnyak03,heinrich03} and nonlocal \cite{kimura06} geometries have also been observed. 
It should be emphasized that these couplings are strongly restricted by the characteristic length scales. 
For example, the dipole coupling decays according to the inverse cube detection law, 
whereas the spin transfer effect by a spin-polarized electric or pure spin current occurs in a system smaller than the spin diffusion length. 


Recently, physical phenomena, such as the spin torque \cite{liu12,liu12c,pai12,kim12,haney13,lee13,garello14,cubukcu14} 
and magnetoresistance effects \cite{huang12,nakayama13,althammer13,hahn13,chen13,liu15,avci15,cho15,kim16}, 
due to the spin Hall effect \cite{dyakonov71,hirsch99,kato04} 
in bilayers consisting of an insulating or metallic ferromagnet and a nonmagnetic heavy metal have attracted much attention. 
The latter, known as the spin Hall magnetoresistance, originates from 
the charge-spin conversion of an external electric current by the direct and inverse spin Hall effects, 
and has been observed by measuring the longitudinal and transverse electric currents, 
which are given by 
\begin{equation}
  \frac{J_{{\rm c}x}}{J_{0}}
  =
  1
  +
  \chi^{\prime\prime}
  +
  \chi
  m_{y}^{2},
  \label{eq:longitudinal_current}
\end{equation}
\begin{equation}
  \frac{J_{{\rm c}y}}{J_{0}}
  =
  -\chi
  m_{x}
  m_{y}
  -
  \chi^{\prime}
  m_{z},
  \label{eq:transverse_current}
\end{equation}
respectively, 
where $J_{0}$ is the electric current density generated by the external electric field. 
The definitions of the dimensionless coefficients, $\chi$, $\chi^{\prime}$, and $\chi^{\prime\prime}$, are given below. 
It should be emphasized that the currents given by Eqs. (\ref{eq:longitudinal_current}) and (\ref{eq:transverse_current}) depend on 
the magnetization direction $\mathbf{m}=(m_{x},m_{y},m_{z})$. 
When another ferromagnet is placed onto the same nonmagnet, 
these currents will be converted to spin current by the spin Hall effect again, and excite spin torque on this additional ferromagnet. 
Then, the magnetization dynamics of two ferromagnets will be coupled through the angular dependencies of the electric current 
given by Eqs. (\ref{eq:longitudinal_current}) and (\ref{eq:transverse_current}). 
This coupling is unavoidable whenever several ferromagnets are placed onto the same nonmagnet, 
and is not restricted by the distance between the ferromagnets because it is carried by the electric current. 
Since the structure consisting of several ferromagnets on the nonmagnetic heavy metal will be important from 
the viewpoints of both fundamental physics and practical applications 
based on the spin Hall effect, such as magnetic random access memory, spin torque oscillators, and bio-inspired computing \cite{locatelli14,grollier16}, 
it is of interest to clarify the role of this coupling. 


In this paper, we investigate the coupled dynamics of magnetizations in ferromagnets in the presence of the spin Hall effect 
by solving the Landau-Lifshitz-Gilbert (LLG) equation numerically for both longitudinal and transverse geometries. 
In addition to the external electric current, the current contributing to the spin Hall magnetoresistance also excites the spin torque. 
The strength of this additional torque is estimated from the theory of the spin Hall magnetoresistance 
extended to the system consisting of several ferromagnets. 
The conventional spin torque is proportional to the spin Hall angle $\vartheta$, 
whereas the new torque is on the order of $\vartheta^{3}$, 
and therefore, its value is two orders of magnitude smaller than the conventional spin torque. 
Nevertheless, it is found that this additional new torque affects the phase difference of the magnetization 
in the self-oscillation state. 
The numerical simulation reveals that the in-phase or antiphase synchronization is observed in the transverse geometry. 
On the other hand, in addition to them, 
the phase difference becomes a quarter of a period in the longitudinal geometry. 
It is found that these phase differences depend on the values of 
the Gilbert damping constant and the dimensionless field-like torque strength. 
An analytical theory clarifying the relation among the current, frequency, and phase difference is also developed. 


The paper is organized as follows. 
In Sec. \ref{sec:System description and LLG equation}, we describe the system under consideration, 
and discuss the theoretical formula of the spin torque excited by the spin Hall effect 
in the presence of the several ferromagnets and the spin Hall magnetoresistance effect. 
In Sec. \ref{sec:Numerical analysis of synchronization}, 
we study the phase differences in the synchronized state of the magnetizations for both the longitudinal and transverse geometries 
by solving the LLG equation numerically. 
In Sec. \ref{sec:Theoretical analysis of current-frequency relation}, 
the theory clarifying the relation among the current, frequency, and phase difference is developed 
based on the LLG equation averaged over constant energy curves. 
The summary of the paper is given in Sec. \ref{sec:Conclusion}. 




\begin{figure}
\centerline{\includegraphics[width=1.0\columnwidth]{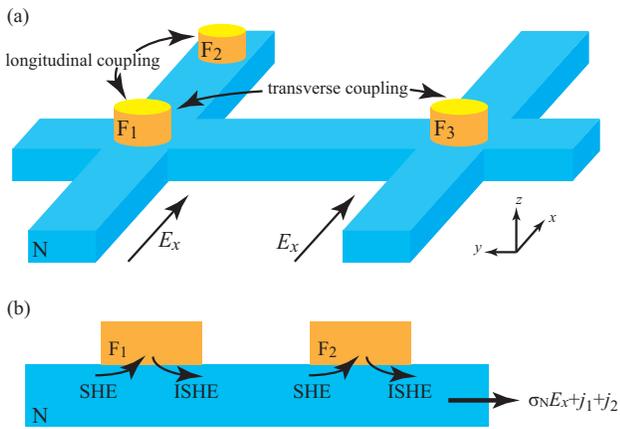}}
\caption{
         (a) Schematic view of the system in this study. 
             Three ferromagnets F${}_{\ell}$ ($\ell=1,2,3$) are placed onto the same nonmagnet N. 
             The external electric field is applied to the $x$ direction. 
         (b) Schematic view of the generations of electric currents by the direct and inverse spin Hall effect (SHE and ISHE) 
             in the longitudinal geometry. 
             The total electric current is the sum of the conventional electric current $J_{0}=\sigma_{\rm N}E_{x}$ 
             and the current generated near the F${}_{1}$/N and F${}_{2}$/N interfaces, $j_{1}$ and $j_{2}$. 
         \vspace{-3ex}}
\label{fig:fig1}
\end{figure}




\section{System description and LLG equation}
\label{sec:System description and LLG equation}

In this section, we describe the system adopted in this study, 
and show the spin torque formulas including the coupling torques between the ferromagnets. 


\subsection{System description}
\label{sec:System description}

The system we consider is schematically shown in Fig. \ref{fig:fig1}(a), 
where three ferromagnets F${}_{\ell}$ ($\ell=1,2,3$) are placed onto the same nonmagnet N. 
We assume that the material parameters in the ferromagnets are identical, for simplicity. 
The external electric field, $E_{x}$, is applied to the $x$ direction, inducing the electric current density $J_{0}=\sigma_{\rm N}E_{x}$, 
where $\sigma_{\rm N}$ is the conductivity of the nonmagnet. 
The direct and inverse spin Hall effects produce electric currents in the longitudinal ($x$) and transverse ($y$) directions. 
These electric currents are converted to the spin current and injected into the F${}_{2}$ and F${}_{3}$ layers due to the spin Hall effect, 
resulting in the excitation of the spin torque. 
Then, the magnetization dynamics in the F${}_{\ell}$ ($\ell=2,3$) layer is affected by that in the F${}_{1}$ layer, and vice versa. 
We call the coupling between the F${}_{1}$ and F${}_{2}$ layers the longitudinal coupling, 
whereas that between the F${}_{1}$ and F${}_{3}$ layers the transverse coupling. 
We assume that both the ferromagnet and nonmagnet are metallic 
because metallic bilayers are generally used to measure the magnetization switching and oscillation by the spin Hall effect \cite{liu12,liu12c,pai12,kim12,garello14}. 
Although the spin Hall magnetoresistance was originally studied for insulating ferromagnets \cite{huang12,nakayama13,althammer13,hahn13}, 
a large spin Hall magnetoresistance in metallic system has also been reported recently \cite{avci15,cho15,kim16}.  


The dimensionless coefficients, $\chi$, $\chi^{\prime}$, and $\chi^{\prime\prime}$, in Eqs. (\ref{eq:longitudinal_current}) and (\ref{eq:transverse_current}) 
for single ferromagnets have been derived for an insulating \cite{chen13} or metallic \cite{kim16,taniguchi16PRB} ferromagnet, 
which are given by 
\begin{equation}
\begin{split}
  \chi
  =&
  \frac{\vartheta^{2}\lambda_{\rm N}}{d_{\rm N}}
  \left[
    {\rm Re}
    \frac{g^{\uparrow\downarrow}}{g_{\rm N}+g^{\uparrow\downarrow} \coth \left( d_{\rm N}/\lambda_{\rm N} \right)}
    -
    \frac{g^{*}}{g_{\rm N}}
  \right]
  \tanh^{2}
  \left(
    \frac{d_{\rm N}}{2 \lambda_{\rm N}}
  \right), 
  \label{eq:chi}
\end{split}
\end{equation}
\begin{equation}
\begin{split}
  \chi^{\prime}
  =
  -\frac{\vartheta^{2} \lambda_{\rm N}}{d_{\rm N}}
  {\rm Im}
  \frac{g^{\uparrow\downarrow}}{g_{\rm N}+g^{\uparrow\downarrow} \coth \left( d_{\rm N}/\lambda_{\rm N} \right)}
  \tanh^{2}
  \left(
    \frac{d_{\rm N}}{2 \lambda_{\rm N}}
  \right), 
  \label{eq:chi_prime}
\end{split}
\end{equation}
\begin{equation}
\begin{split}
  \chi^{\prime\prime}
  =&
  \frac{2 \vartheta^{2} \lambda_{\rm N}}{d_{\rm N}}
  \tanh
  \left(
    \frac{d_{\rm N}}{2 \lambda_{\rm N}}
  \right)
\\
  &-
  \frac{\vartheta^{2} \lambda_{\rm N}}{d_{\rm N}}
  {\rm Re}
  \frac{g^{\uparrow\downarrow} }{g_{\rm N} + g^{\uparrow\downarrow} \coth \left( d_{\rm N}/\lambda_{\rm N} \right)}
  \tanh^{2}
  \left(
    \frac{d_{\rm N}}{2 \lambda_{\rm N}}
  \right),
  \label{eq:chi_prime_prime}
\end{split}
\end{equation}
where $d_{\rm N}$ and $\lambda_{\rm N}$ are 
thickness and spin diffusion length of the nonmagnet, respectively, 
whereas $g_{\rm N}/S=h \sigma_{\rm N}/\left( 2e^{2} \lambda_{\rm N} \right)$ with the cross section area of the ferromagnetic/nonmagnetic interface $S$. 
The dimensionless mixing conductance $g^{\uparrow\downarrow}=g_{\rm r}^{\uparrow\downarrow}+ig_{\rm i}^{\uparrow\downarrow}$ consists of 
its real and imaginary parts \cite{brataas00,brataas01,zwierzycki05}, 
and $g^{*}$ is defined as 
\begin{equation}
  \frac{1}{g^{*}}
  =
  \frac{2}{(1-p_{g}^{2})g}
  +
  \frac{1}{g_{\rm F} \tanh \left( d_{\rm F}/\lambda_{\rm F} \right)}
  +
  \frac{1}{g_{\rm N} \tanh \left( d_{\rm N}/\lambda_{\rm N} \right)}. 
\end{equation}
Here, $g=g^{\uparrow\uparrow}+g^{\downarrow\downarrow}$ is the sum of the conductances of the spin-up and spin-down electrons, 
whereas $p_{g}=(g^{\uparrow\uparrow}-g^{\downarrow\downarrow})/g$ is its spin polarization. 
The ferromagnetic/nonmagnetic interface resistance $r$ is related to $g$ via $g/S=(h/e^{2})/r$. 
We also introduce $g_{\rm F}/S=h \left( 1 - p_{\sigma}^{2} \right)\sigma_{\rm F}/\left( 2e^{2} \lambda_{\rm F} \right)$, 
where $\sigma_{\rm F}$ is the conductivity of the ferromagnet and $p_{\sigma}$ is its spin polarization. 
The thickness and spin diffusion length of the ferromagnet are denoted as $d_{\rm F}$ and $\lambda_{\rm F}$, respectively. 
The term related to $g^{*}$ is neglected when the ferromagnet is an insulator \cite{chen13}, i.e., $r \to \infty$. 
The following quantities correspond to the effective spin polarizations 
of the damping-like (or Slonczewski \cite{slonczewski96}) torque and the field-like torque, respectively, 
\begin{equation}
  \vartheta_{\rm R(I)}
  =
  \vartheta
  \tanh
  \left(
    \frac{d_{\rm N}}{2 \lambda_{\rm N}}
  \right)
  {\rm Re}
  \left(
    {\rm Im}
  \right)
  \frac{g^{\uparrow\downarrow}}{g_{\rm N}+g^{\uparrow\downarrow} \coth \left( d_{\rm N}/\lambda_{\rm N} \right)}. 
  \label{eq:effective_SHA}
\end{equation}
For the later discussion, we introduce 
\begin{equation}
  \beta
  =
  -\frac{\vartheta_{\rm I}}{\vartheta_{\rm R}}, 
  \label{eq:beta}
\end{equation}
which corresponds to the ratio of the field-like torque to the damping-like torque. 


The values of the parameters used in the following calculations are derived from recent experiments on the W/CoFeB heterostructure \cite{kim16}, 
where $\rho_{\rm F}=1/\sigma_{\rm F}=1.6$ k$\Omega$nm, $p_{\sigma}=0.72$, $\lambda_{\rm F}=1.0$ nm, 
$\rho_{\rm N}=1/\sigma_{\rm N}=1.25$ k$\Omega$nm, $\lambda_{\rm N}=1.2$ nm, and $\vartheta=0.27$, 
whereas the thicknesses are assumed to be $d_{\rm F}=2$ nm and $d_{\rm N}=3$ nm. 
The interface conductances were not evaluated in Ref. \cite{kim16} by assuming a transparent interface. 
Instead, we use typical values of the interface conductances obtained from the first-principles calculations \cite{zwierzycki05}, 
$r=0.25$ k$\Omega$nm${}^{2}$, $p_{g}=0.5$, and $g_{\rm r}^{\uparrow\downarrow}/S=25$ nm${}^{-2}$. 
We note that the imaginary part of the mixing conductance, $g_{\rm i}^{\uparrow\downarrow}$, is either positive or negative, 
depending on the material and thickness \cite{zwierzycki05}. 
The sign of $g_{\rm i}^{\uparrow\downarrow}$ determines those of $\chi^{\prime}$ and $\vartheta_{\rm I}$, or equivalently, $\beta$. 
For example, when $g_{\rm i}^{\uparrow\downarrow}/S=1$ nm${}^{-2}$, 
$\chi \simeq 0.010$, $\chi^{\prime} \simeq -0.0002$, 
$\chi^{\prime\prime} \simeq 0.035$, $\theta_{\rm R} \simeq 0.167$, and $\theta_{\rm I} \simeq 0.002$ ($\beta \simeq -0.010$); 
see Appendix \ref{sec:Values of parameters in numerical simulations}. 
In the following calculations, we study the magnetization dynamics for several values of $\beta$. 


\subsection{Spin torques in longitudinal and transverse geometries}

Equation (\ref{eq:longitudinal_current}) was derived for a system having single ferromagnet. 
In this case, $J_{0}$ is the external electric current density, 
whereas $(\chi^{\prime\prime}+\chi m_{y})J_{0}$ is the current density generated 
as a result of the charge-spin conversion by the direct and inverse spin Hall effects. 
In the longitudinal geometry in the present study, on the other hand, 
two ferromagnetic/nonmagnetic interfaces, i.e., F${}_{1}$/N and F${}_{2}$/N interfaces, 
contribute to the generation of the longitudinal current through the direct and inverse spin Hall effects, as schematically shown in Fig. \ref{fig:fig1}(b). 
Let us denote the electric current density generated by these effects near the F${}_{\ell}$/N interface as $j_{\ell x}$ ($\ell=1,2$). 
This current density is determined by the conservation law of the electric current, as follows. 
Considering in a similar manner to the case of the single ferromagnet, 
the electric current $J_{0}+j_{1x}$ is converted to the spin current by the spin Hall effect near the F${}_{2}$/N interface, 
and this spin current produces an additional electric current $(\chi^{\prime\prime}+\chi m_{2y}^{2})(J_{0}+j_{1x})$ by the inverse spin Hall effect. 
Therefore, the total longitudinal electric current density near the F${}_{2}$/N interface is 
$(1+\chi^{\prime\prime}+\chi m_{2y}^{2})(J_{0}+j_{1x})$. 
Similarly, the electric current density near the F${}_{1}$/N interface can be expressed as $(1+\chi^{\prime\prime} + \chi m_{1y}^{2})(J_{0}+j_{2x})$. 
These currents should be equal to the total electric current density, $J_{0}+j_{1x}+j_{2x}$, according to the conservation law of the electric current. 
Then, we find that $j_{\ell x}$ is given by 
\begin{equation}
\begin{split}
  j_{\ell x}
  &=
  \frac{\left( \chi^{\prime\prime} + \chi m_{\ell y}^{2} \right) \left( 1 + \chi^{\prime\prime} + \chi m_{\ell^{\prime} y}^{2} \right)}
    {1 - \left( \chi^{\prime\prime} + \chi^{\prime} m_{\ell y}^{2} \right) \left( \chi^{\prime\prime} + \chi m_{\ell^{\prime} y}^{2} \right)}
  J_{0}
\\
  &\simeq
  \left(
    \chi^{\prime\prime}
    +
    \chi m_{\ell y}^{2}
  \right)
  J_{0},
  \label{eq:electric_current_x}
\end{split}
\end{equation}
where we neglect the higher order terms of $\vartheta$. 
Then, the spin torque excited on the F${}_{\ell}$ layer by the spin Hall effect in the longitudinal geometry 
is obtained by replacing the external electric current density $J_{0}=\sigma_{\rm N}E_{x}$ in the previous work \cite{chen13} 
with $J_{0}+j_{\ell^{\prime}x}\simeq \left( 1 + \chi^{\prime\prime} + \chi m_{\ell^{\prime} y}^{2} \right)J_{0}$, 
and is given by 
\begin{equation}
\begin{split}
  \mathbf{T}_{\ell}^{\rm L}
  =&
  -\frac{\gamma \hbar \vartheta_{\rm R}J_{0}}{2eMd_{\rm F}}
  \left(
    1
    +
    \chi^{\prime\prime}
    +
    \chi
    m_{\ell^{\prime} y}^{2}
  \right)
  \mathbf{m}_{\ell}
  \times
  \left(
    \mathbf{e}_{y}
    \times
    \mathbf{m}_{\ell}
  \right)
\\
  &-
  \frac{\gamma \hbar \vartheta_{\rm I} J_{0}}{2eMd_{\rm F}}
  \left(
    1
    +
    \chi^{\prime\prime}
    +
    \chi
    m_{\ell^{\prime} y}^{2}
  \right)
  \mathbf{e}_{y}
  \times
  \mathbf{m}_{\ell}, 
  \label{eq:torque_L}
\end{split}
\end{equation}
where $(\ell,\ell^{\prime})=(1,2)$ or $(2,1)$. 
The unit vector pointing in the magnetization direction of the F${}_{\ell}$ layer is $\mathbf{m}_{\ell}$, 
and $\gamma$, $M$, and $d$ are the gyromagnetic ratio, 
the saturation magnetization, and the thickness of the ferromagnet, respectively. 


Note that the terms, 
\begin{equation}
\begin{split}
  \mathbf{T}_{\ell}^{(0)}
  =&
  -\frac{\gamma \hbar \vartheta_{\rm R}J_{0}}{2eMd_{\rm F}}
  \mathbf{m}_{\ell}
  \times
  \left(
    \mathbf{e}_{y}
    \times
    \mathbf{m}_{\ell}
  \right)
  -
  \frac{\gamma \hbar \vartheta_{\rm I} J_{0}}{2eMd_{\rm F}}
  \mathbf{e}_{y}
  \times
  \mathbf{m}_{\ell}, 
  \label{eq:torque_conventional}
\end{split}
\end{equation}
in Eq. (\ref{eq:torque_L}) are 
the conventional spin torques generated by the external electric current $J_{0}$ and the spin Hall effect, 
and are often referred to as the damping-like torque and the field-like torque, respectively. 
On the other hand, the terms 
\begin{equation}
\begin{split}
  &
  -\frac{\gamma \hbar \vartheta_{\rm R}J_{0}}{2eMd_{\rm F}}
  \left(
    \chi^{\prime\prime}
    +
    \chi
    m_{\ell^{\prime} y}^{2}
  \right)
  \mathbf{m}_{\ell}
  \times
  \left(
    \mathbf{e}_{y}
    \times
    \mathbf{m}_{\ell}
  \right)
\\
  &-
  \frac{\gamma \hbar \vartheta_{\rm I} J_{0}}{2eMd_{\rm F}}
  \left(
    \chi^{\prime\prime}
    +
    \chi
    m_{\ell^{\prime} y}^{2}
  \right)
  \mathbf{e}_{y}
  \times
  \mathbf{m}_{\ell}, 
  \label{eq:torque_L_coupling}
\end{split}
\end{equation}
in Eq. (\ref{eq:torque_L}) originate from the current $j_{\ell^{\prime}x}$, 
and are newly introduced in this study. 
It should be emphasized that Eq. (\ref{eq:torque_L_coupling}) 
depends on the magnetization direction of the other ferromagnet F${}_{\ell^{\prime}}$ ($\ell^{\prime}=1$ or $2$), $\mathbf{m}_{\ell^{\prime}}$, 
resulting in the coupling between the F${}_{1}$ and F${}_{2}$ layers. 


Let us next show the spin torque formulas in the transverse geometry. 
We denote the electric current density flowing in the $y$ direction 
generated near the F${}_{\ell}$/N ($\ell=1,3$) interface by the inverse spin Hall effect as $j_{\ell y}$. 
The conservation law of the total electric current density, $j_{1y}+j_{3y}$, gives 
(see also Appendix \ref{sec:Derivation of coupling torque in transverse geometry})
\begin{equation}
\begin{split}
  j_{\ell y}
  &=
  -\left[
    \frac{\left( \chi m_{\ell x} m_{\ell y} + \chi^{\prime} m_{\ell z} \right)}
      {1 - \left( \chi^{\prime\prime} + \chi m_{\ell x}^{2} \right) \left( \chi^{\prime\prime} + \chi m_{\ell^{\prime} x}^{2} \right)}
  \right.
\\
  &
  \left.
  \ \ \ \ + 
    \frac{\left( \chi^{\prime\prime} + \chi m_{\ell x}^{2} \right) \left( \chi m_{\ell^{\prime} x} m_{\ell^{\prime} y} + \chi^{\prime} m_{\ell^{\prime} z} \right)}
      {1 - \left( \chi^{\prime\prime} + \chi m_{\ell x}^{2} \right) \left( \chi^{\prime\prime} + \chi m_{\ell^{\prime} x}^{2} \right)}
  \right]
  J_{0}
\\
  &\simeq
  -\left(
    \chi
    m_{\ell x}
    m_{\ell y}
    +
    \chi^{\prime}
    m_{\ell z}
  \right)
  J_{0}. 
  \label{eq:electric_current_y}
\end{split}
\end{equation}
In addition to the conventional spin torque, Eq. (\ref{eq:torque_conventional}), 
this transverse electric current also excites the spin torque on the other ferromagnet. 
In total, the spin torque acting on the F${}_{\ell}$ layer is given by 
[$(\ell,\ell^{\prime})=(1,3)$ or $(3,1)$] 
\begin{equation}
\begin{split}
  \mathbf{T}_{\ell}^{\rm T}
  =&
  -\frac{\gamma \hbar \vartheta_{\rm R} J_{0}}{2e Md_{\rm F}}
  \mathbf{m}_{\ell}
  \times
  \left(
    \mathbf{e}_{y}
    \times
    \mathbf{m}_{\ell}
  \right)
  -
  \frac{\gamma \hbar \vartheta_{\rm I} J_{0}}{2e Md_{\rm F}}
  \mathbf{e}_{y}
  \times
  \mathbf{m}_{\ell}
\\
  &
  -
  \frac{\gamma \hbar \vartheta_{\rm R} \left( \chi m_{\ell^{\prime} x} m_{\ell^{\prime} y} + \chi^{\prime} m_{\ell^{\prime} z} \right) J_{0}}{2eMd_{\rm F}}
  \mathbf{m}_{\ell}
  \times
  \left(
    \mathbf{e}_{x}
    \times
    \mathbf{m}_{\ell}
  \right)
\\
  &-
  \frac{\gamma \hbar \vartheta_{\rm I} \left( \chi m_{\ell^{\prime} x} m_{\ell^{\prime} y} + \chi^{\prime} m_{\ell^{\prime} z} \right) J_{0}}{2eMd_{\rm F}}
  \mathbf{e}_{x}
  \times
  \mathbf{m}_{\ell},
  \label{eq:torque_T}
\end{split}
\end{equation}
The last two terms represent the coupling torque between the F${}_{1}$ and F${}_{3}$ layers. 
Note that the direction of this coupling torque is different from that of the conventional torque 
because the currents $J_{0}$ and $j_{\ell y}$ flow in different directions. 


In the following, the torques related to $\chi$, $\chi^{\prime}$, and $\chi^{\prime\prime}$ 
in Eqs. (\ref{eq:torque_L}) and (\ref{eq:torque_T}) are referred to as coupling torque. 
The ratio of these new torques to the conventional spin torque is on the order of $\chi \propto \vartheta^{2} \sim 10^{-2}$. 
Since the conventional spin torque due is proportional to the spin Hall angle $\vartheta_{\rm R} \propto \vartheta$, 
the coupling torque is proportional to $\vartheta^{3}$. 
Although the spin Hall angle is usually a small quantity, 
it will be shown that the coupling torques play a non-negligible role 
on the magnetization dynamics, as shown below. 


\subsection{LLG equation}
\label{sec:LLG equation}

In the following sections, we study the magnetization dynamics excited by the spin torque given by Eq. (\ref{eq:torque_L}) or (\ref{eq:torque_T}) 
both numerically and analytically. 
We neglect the transverse coupling when the role of the longitudinal coupling is studied, and vice versa, for simplicity, 
which corresponds to considering a system consisting of the F${}_{1}$ and F${}_{2}$ layers, or F${}_{1}$ and F${}_{3}$ layers. 
The magnetization dynamics in the F${}_{\ell}$ ($\ell=1,2,3$) is described by the LLG equation, 
\begin{equation}
\begin{split}
  \frac{d \mathbf{m}_{\ell}}{dt}
  =&
  -\gamma
  \mathbf{m}_{\ell}
  \times
  \mathbf{H}_{\ell}
  +
  \alpha
  \mathbf{m}_{\ell}
  \times
  \frac{d \mathbf{m}_{\ell}}{dt}
  +
  \mathbf{T}_{\ell}^{\rm L,T},
  \label{eq:LLG}
\end{split}
\end{equation}
where the Gilbert damping constant is denoted as $\alpha$. 
In the following calculations, 
we use the values of the material parameters, 
$\gamma=1.764 \times 10^{7}$ rad/(Oe s) and $\alpha=0.005$, 
derived from the experiments \cite{tsunegi14}. 
For the later discussion, it is useful to show the explicit forms of the LLG equation in 
the longitudinal and transverse geometries. 
Equation (\ref{eq:LLG}) for the longitudinal geometry is
\begin{equation}
\begin{split}
  &
  \left(
    1
    +
    \alpha^{2}
  \right)
  \frac{d \mathbf{m}_{\ell}}{d t}
  =
  -\gamma
  \mathbf{m}_{\ell}
  \times
  \mathbf{H}_{\ell}
\\
  &-
  \gamma
  \left(
    1
    +
    \chi^{\prime\prime}
  \right)
  \frac{\hbar \vartheta_{\rm R} J_{0}}{2eMd_{\rm F}}
  \mathbf{m}_{\ell}
  \times
  \left(
    \mathbf{e}_{y}
    \times
    \mathbf{m}_{\ell}
  \right)
  -
  \alpha
  \gamma
  \mathbf{m}_{\ell}
  \times
  \left(
    \mathbf{m}_{\ell}
    \times
    \mathbf{H}_{\ell}
  \right)
\\
  &-
  \gamma
  \chi
  m_{\ell^{\prime} y}^{2}
  \frac{\hbar \vartheta_{\rm R} J_{0}}{2eMd_{\rm F}}
  \mathbf{m}_{\ell}
  \times
  \left(
    \mathbf{e}_{y}
    \times
    \mathbf{m}_{\ell}
  \right)
\\
  &-
  \gamma
  \left(
    1
    +
    \chi^{\prime\prime}
  \right)
  \left(
    \alpha
    +
    \beta
  \right)
  \frac{\hbar \vartheta_{\rm R} J_{0}}{2eMd_{\rm F}}
  \mathbf{m}_{\ell}
  \times
  \mathbf{e}_{y}
\\
  &
  +
  \gamma
  \left(
    1
    +
    \chi^{\prime\prime}
  \right)
  \alpha
  \beta
  \frac{\hbar \vartheta_{\rm R} J_{0}}{2eMd_{\rm F}}
  \mathbf{m}_{\ell}
  \times
  \left(
    \mathbf{e}_{y}
    \times
    \mathbf{m}_{\ell}
  \right)
\\
  &-
  \gamma
  \left(
    \alpha
    +
    \beta
  \right)
  \chi 
  m_{\ell^{\prime} y}^{2}
  \frac{\hbar \vartheta_{\rm R} J_{0}}{2eMd_{\rm F}}
  \mathbf{m}_{\ell}
  \times
  \mathbf{e}_{y}
\\
  &+
  \gamma
  \alpha
  \beta 
  \chi
  m_{\ell^{\prime} y}^{2}
  \frac{\hbar \vartheta_{\rm R} J_{0}}{2eMd_{\rm F}}
  \mathbf{m}_{\ell}
  \times
  \left(
    \mathbf{e}_{y}
    \times
    \mathbf{m}_{\ell}
  \right), 
  \label{eq:LLG_L_exact}
\end{split}
\end{equation}
whereas that for the transverse geometry is 
\begin{equation}
\begin{split}
  &
  \left(
    1
    +
    \alpha^{2}
  \right)
  \frac{d \mathbf{m}_{\ell}}{dt}
  =
  -\gamma
  \mathbf{m}_{\ell}
  \times
  \mathbf{H}_{\ell}
\\
  &-
  \gamma
  \frac{\hbar \vartheta_{\rm R} J_{0}}{2eMd_{\rm F}}
  \mathbf{m}_{\ell}
  \times
  \left(
    \mathbf{e}_{y}
    \times
    \mathbf{m}_{\ell}
  \right)
  -
  \alpha
  \gamma
  \mathbf{m}_{\ell}
  \times
  \left(
    \mathbf{m}_{\ell}
    \times
    \mathbf{H}_{\ell}
  \right)
\\
  &-
  \gamma
  \left(
    \chi
    m_{\ell^{\prime} x}
    m_{\ell^{\prime} y}
    +
    \chi^{\prime}
    m_{\ell^{\prime} z}
  \right)
  \frac{\hbar \vartheta_{\rm R} J_{0}}{2eMd_{\rm F}}
  \mathbf{m}_{\ell}
  \times
  \left(
    \mathbf{e}_{x}
    \times
    \mathbf{m}_{\ell}
  \right)
\\
  &-
  \gamma
  \left(
    \alpha
    +
    \beta
  \right)
  \frac{\hbar \vartheta_{\rm R} J_{0}}{2eMd_{\rm F}}
  \mathbf{m}_{\ell}
  \times
  \mathbf{e}_{y}
\\
  &
  +
  \gamma
  \alpha
  \beta
  \frac{\hbar \vartheta_{\rm R} J_{0}}{2eMd_{\rm F}}
  \mathbf{m}_{\ell}
  \times
  \left(
    \mathbf{e}_{y}
    \times
    \mathbf{m}_{\ell}
  \right)
\\
  &-
  \gamma
  \left(
    \alpha
    +
    \beta
  \right)
  \left(
    \chi
    m_{\ell^{\prime} x}
    m_{\ell^{\prime} y}
    +
    \chi^{\prime}
    m_{\ell^{\prime} z}
  \right)
  \frac{\hbar \vartheta_{\rm R} J_{0}}{2eMd_{\rm F}}
  \mathbf{m}_{\ell}
  \times
  \mathbf{e}_{x}
\\
  &+
  \gamma
  \alpha
  \beta 
  \left(
    \chi
    m_{\ell^{\prime} x}
    m_{\ell^{\prime} y}
    +
    \chi^{\prime}
    m_{\ell^{\prime} z}
  \right)
  \frac{\hbar \vartheta_{\rm R} J_{0}}{2eMd_{\rm F}}
  \mathbf{m}_{\ell}
  \times
  \left(
    \mathbf{e}_{x}
    \times
    \mathbf{m}_{\ell}
  \right). 
  \label{eq:LLG_T_exact}
\end{split}
\end{equation}






\section{Numerical analysis of synchronization}
\label{sec:Numerical analysis of synchronization}

In this section, we study the magnetization dynamics in the ferromagnets 
in the presence of the coupling torques by solving Eq. (\ref{eq:LLG}) numerically. 
The self-oscillation of the magnetization provides 
an interesting example to understand the role of the coupling torque. 
Note that the coupling torques are proportional to the parameters $\chi$, $\chi^{\prime}$, and $\chi^{\prime\prime}$, 
and their products to other parameters such as $\alpha\beta\chi$, the values of which are relatively small, 
as can be seen in Eqs. (\ref{eq:LLG_L_exact}) and (\ref{eq:LLG_T_exact}). 
Nevertheless, the coupling torques lead to the phase synchronization, as shown below. 


The self-oscillation of the magnetization in single ferromagnets by the spin Hall effect has been observed 
for in-plane magnetized ferromagnets \cite{liu12c}, 
and is induced by the conventional spin torque given by Eq. (\ref{eq:torque_conventional}). 
Therefore, in the following, we assume that the magnetic field, 
\begin{equation}
  \mathbf{H}_{\ell}
  =
  H_{\rm K}
  m_{\ell y}
  \mathbf{e}_{y}
  -
  4 \pi M 
  m_{\ell z}
  \mathbf{e}_{z}, 
  \label{eq:field}
\end{equation}
consists of an in-plane anisotropy field $H_{\rm K}$ along the $y$ direction 
and a demagnetization field $4\pi M$ in the $z$ direction, 
which we assume as $H_{\rm K}=200$ Oe and $M=1500$ emu/c.c. \cite{kim16} in the following calculations. 
It is known for the case of the single ferromagnet \cite{hillebrands06} that 
the in-plane self-oscillation can be excited when 
the electric current density $J_{0}$ is in the range of $J_{\rm c} < J_{0} < J^{*}$, where 
\begin{equation}
  J_{\rm c}
  =
  \frac{2 \alpha eMd_{\rm F}}{\hbar \vartheta_{\rm R}}
  \left(
    H_{\rm K}
    +
    2 
    \pi 
    M
  \right), 
  \label{eq:J_c_single}
\end{equation}
\begin{equation}
  J^{*}
  =
  \frac{4 \alpha eMd_{\rm F}}{\pi \hbar \vartheta_{\rm R}}
  \sqrt{4\pi M \left( H_{\rm K} + 4 \pi M \right)}, 
  \label{eq:J_switch_single}
\end{equation}
which in this study are $J_{\rm c} \simeq 26$ and $J^{*} \simeq 33$ MA/cm${}^{2}$. 




\begin{figure}
\centerline{\includegraphics[width=1.0\columnwidth]{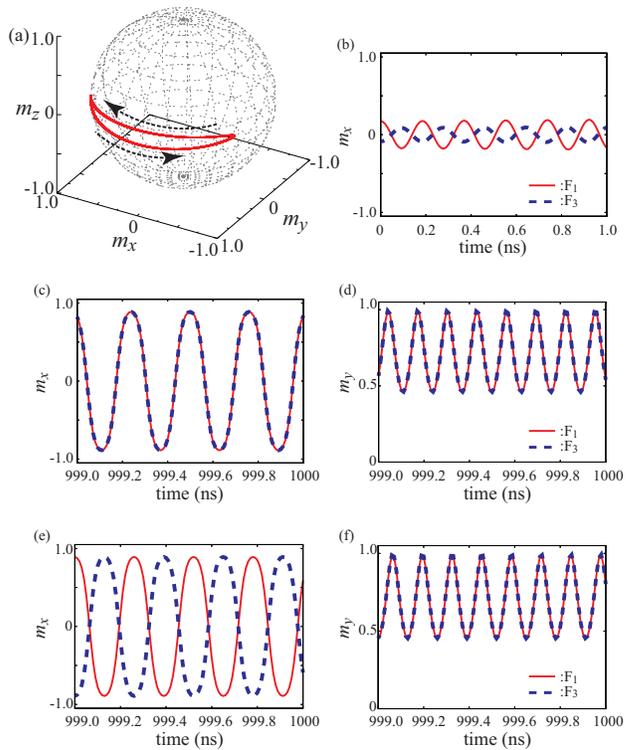}}
\caption{
         (a) A typical in-plane self-oscillation trajectory of the magnetization obtained by solving Eq. (\ref{eq:LLG}) numerically for the transverse geometry. 
         The dotted lines indicate the oscillation direction. 
         (b) The time evolutions of $m_{1x}(t)$ and $m_{1y}(t)$ near the initial state. 
         The time variations of the $x$ and $y$ components of the magnetizations in the steady state are 
         shown in (c) and (d) for $\beta = -0.01$ and (e) and (f) for $\beta = +0.01$. 
         The solid red lines correspond to time evolution in F${}_{1}$, 
         and dotted blue lines to those in F${}_{3}$ for (b) through (f). 
         \vspace{-3ex}}
\label{fig:fig2}
\end{figure}




\subsection{Transverse geometry}

Let us first investigate the magnetization dynamics in the transverse geometry 
because this geometry provides a simple example of the coupled motion. 
We start with solving the LLG equation given by Eq. (\ref{eq:LLG_T_exact}) for the F${}_{1}$ and F${}_{3}$ layers. 
Figure \ref{fig:fig2}(a) shows an example of the trajectory of the magnetization dynamics obtained by solving Eq. (\ref{eq:LLG_T_exact}) numerically, 
where $J_{0}=30$ MA/cm${}^{2}$. 
As shown, the in-plane oscillation is observed in the steady state. 
The initial conditions set for $\mathbf{m}_{1}$ and $\mathbf{m}_{3}$ are different as 
$\mathbf{m}_{1}(0)=(\cos 80^{\circ},\sin 80^{\circ},0)$ and $\mathbf{m}_{3}(0)=(\cos 95^{\circ},\sin 95^{\circ},0)$. 
Therefore, the dynamics of $\mathbf{m}_{1}$ and $\mathbf{m}_{3}$ near the initial time are different, as shown in Fig. \ref{fig:fig2}(b), 
where the time evolutions of $m_{1x}(t)$ and $m_{3x}(t)$ in $0 \le t \le 1$ ns are shown. 
Nevertheless, the dynamics of $\mathbf{m}_{1}$ and $\mathbf{m}_{3}$ synchronize gradually, 
and finally, synchronization of $m_{1x}(t)=m_{3x}(t)$ and $m_{1y}(t)=m_{3y}(t)$ is realized, 
as shown in Figs. \ref{fig:fig2}(c) and \ref{fig:fig2}(d). 
We emphasize here that the dynamics shown in these figures are obtained for $\beta = -0.01$. 


The mutual, as well as self, synchronization of the spin torque induced magnetization oscillation 
by using spin waves, electric current, microwave field, or dipole coupling has been an exciting topic 
from the viewpoints of both nonlinear science and practical applications 
\cite{kaka05,mancoff05,grollier06,slavin06,persson07,ruotolo09,georges08,zhou09,urazhdin10,sani13,demidov14,locatelli15,khalsa15,kendziorczyk16,tsunegi16,awad17,kudo17}. 
The key quantity of the synchronization is the phase difference $\Delta\varphi$ between each magnetization 
to enhance the emission power from the spin torque oscillators 
and to investigate new practical applications such as neuromorphic architectures \cite{locatelli14,grollier16}. 
The synchronization found here, i.e., $m_{\ell x}(t)=m_{\ell^{\prime} x}(t)$, is called the in-phase synchronization. 
We should emphasize here that, although the results shown in Figs. \ref{fig:fig2}(b) and \ref{fig:fig2}(c) are shown for one certain initial condition, 
the in-phase synchronizations are confirmed for the present set of the parameters even when the initial conditions are changed. 




\begin{figure}
\centerline{\includegraphics[width=1.0\columnwidth]{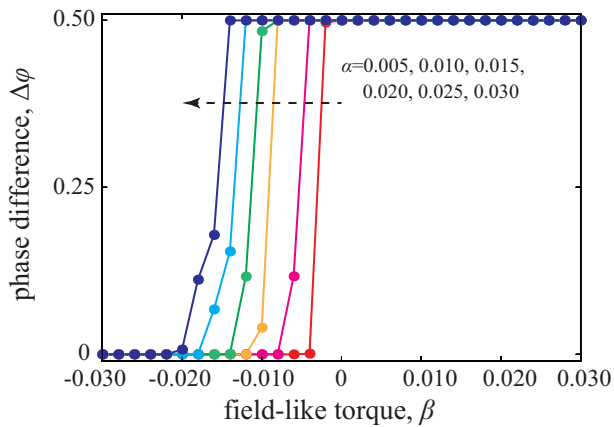}}
\caption{
         Dependences of the phase differences, $\Delta\varphi$, for several values of $\alpha$ on the field-like torque strength $\beta$ in the transverse geometry. 
         $\Delta\varphi=0$ and $0.5$ correspond to the in-phase and antiphase, respectively. 
         The values of the current density, $J_{0}$, is increased linearly, where $J_{0}=30$ MA/cm${}^{2}$ for $\alpha=0.005$. 
         \vspace{-3ex}}
\label{fig:fig3}
\end{figure}




On the other hand, it was shown for the case of the current-injection locking of the spin torque oscillator that 
the phase difference between the magnetization oscillation and the alternative current depends on the strength of the field-like torque \cite{zhou09}. 
The field-like torque in the present system can be either positive or negative, as mentioned above. 
These facts motivate us to study the magnetization dynamics for different values of $\beta$. 
When $\beta=+0.01$, synchronized dynamics is observed in a similar manner, but in this case, 
the phase difference is antiphase, i.e., $m_{1x}(t)=-m_{3x}(t)$, 
as shown in Figs. \ref{fig:fig2}(e) and \ref{fig:fig2}(f). 


Figure \ref{fig:fig3} summarizes the dependences of the phase difference, $\Delta\varphi$, between the magnetizations 
on the field-like torque strength $\beta$ for the several values of the damping constant $\alpha$. 
The vertical axis in this figure represents the phase difference in the unit of the oscillation period; 
i.e., $\Delta\varphi=0$ corresponds to the in-phase synchronization. 
On the other hand, $\Delta\varphi=0.50$ means that the phase difference is half of a period, 
and thus, is antiphase. 
The algorithm evaluating $\Delta\varphi$ in the numerical simulation is summarized in Appendix \ref{sec:Values of parameters in numerical simulations}. 
The results indicate that the phase difference is antiphase for positive $\beta$, 
whereas it becomes in-phase when $\beta$ becomes smaller than a certain value, 
except for the narrow intermediate region near $\beta \sim -\alpha/2$, where the phase difference is in between in-phase and antiphase.




\begin{figure}
\centerline{\includegraphics[width=1.0\columnwidth]{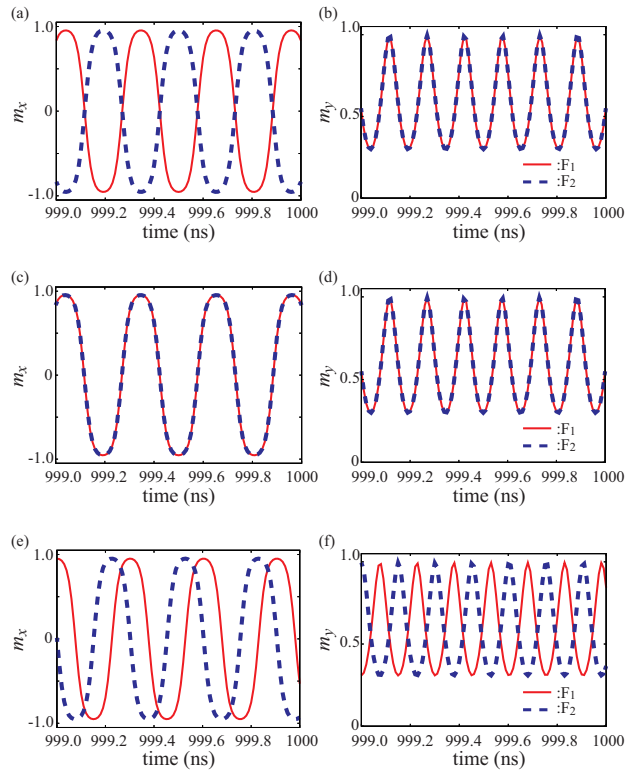}}
\caption{
         The variations of the $x$ and $y$ components of the magnetizations in the F${}_{1}$ (solid, red) and F${}_{2}$ (dotted, blue) in the longitudinal geometry are 
         shown in (a), (c) and (b), (d) respectively, 
         where $\beta = -0.01$. 
         The initial conditions of $\mathbf{m}_{2}$ are different between (a), (b) and (c), (d). 
         The magnetization dynamics for $\beta = +0.01$ are shown in (e) and (f) 
         \vspace{-3ex}}
\label{fig:fig4}
\end{figure}




\subsection{Longitudinal geometry}

Next, we study the magnetization dynamics in the longitudinal geometry between F${}_{1}$ and F${}_{2}$ layers by solving Eq. (\ref{eq:LLG_L_exact}) numerically. 
Figure \ref{fig:fig4}(a) and \ref{fig:fig4}(b) show $m_{\ell x}(t)$ and $m_{\ell y}(t)$ in a steady state, 
where $\beta = -0.01$. 
The initial conditions in these figures are $\mathbf{m}_{1}(0)=(\cos 80^{\circ},\sin 80^{\circ},0)$ and $\mathbf{m}_{2}(0)=(\cos 95^{\circ},\sin 95^{\circ},0)$. 
An antiphase synchronization is observed in this case. 
We notice, however, that the in-phase synchronization can also be realized when the initial conditions are changed. 
Figure \ref{fig:fig4}(c) and \ref{fig:fig4}(d) show such an example, 
where $\mathbf{m}_{1}(0)=(\cos 80^{\circ},\sin 80^{\circ},0)$ and $\mathbf{m}_{2}(0)=(\cos 85^{\circ},\sin 85^{\circ},0)$ are assumed. 
These numerical results indicate that both the in-phase and antiphase synchronizations are stable in this case, 
and whether the phase difference finally becomes in-phase or antiphase depends on the initial conditions, material parameters, and current magnitude. 
We also notice that the phase difference is changed when the value of $\beta$ is changed. 
Figures \ref{fig:fig4}(e) and \ref{fig:fig4}(f) show $m_{\ell x}$ and $m_{\ell y}$ for $\beta = +0.01$. 
In this case, the phase difference of the magnetizations is a quarter of a precession period. 




\begin{figure}
\centerline{\includegraphics[width=1.0\columnwidth]{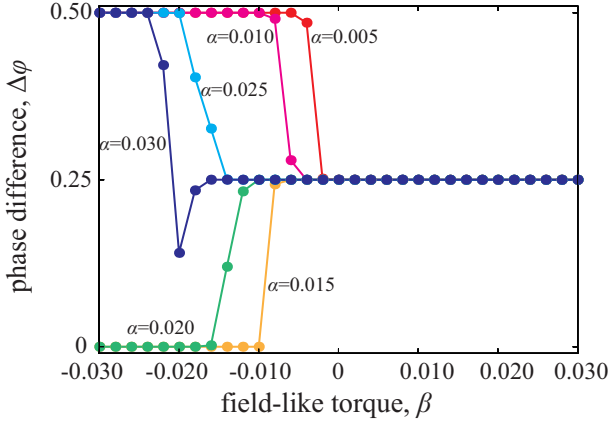}}
\caption{
         Dependences of the phase differences, $\Delta\varphi$, for several values of $\alpha$ on the field-like torque strength $\beta$ in the longitudinal geometry, 
         where $\Delta\varphi=0$ and $0.5$ correspond to the in-phase and antiphase, respectively.
         $\Delta\varphi=0.25$ means that the phase difference is a quarter of a period. 
         The values of the current density, $J_{0}$, is increased linearly, where $J_{0}=30$ MA/cm${}^{2}$ for $\alpha=0.005$. 
         \vspace{-3ex}}
\label{fig:fig5}
\end{figure}




The dependences of the phase difference on the field-like torque strength $\beta$ for several values of the Gilbert damping constant $\alpha$ 
are summarized in Fig. \ref{fig:fig5}, 
where $\Delta\varphi=0.25$ in this figure means that the phase difference is a quarter of a period. 
The phase difference is found to become a quarter of the period for positive $\beta$, 
whereas it becomes in-phase or antiphase for negative $\beta$, depending on the initial states of the magnetizations. 



\subsection{Summary of numerical simulations}

Let us summarize the results of the numerical simulations here. 
In the transverse geometry, the coupling torque induces the synchronized oscillation of the magnetizations 
and finally stabilizes the configuration in the in-phase or antiphase state, 
depending on the values of the field-like torque strength $\beta$ and the damping constant $\alpha$. 
The phase difference in the stable synchronized state in the longitudinal geometry also depends on the values of $\beta$ and $\alpha$, 
as well as the initial conditions. 
In this case, however, in addition to the in-phase or antiphase state, 
a phase difference with a quarter of a period is generated.




\begin{figure}
\centerline{\includegraphics[width=1.0\columnwidth]{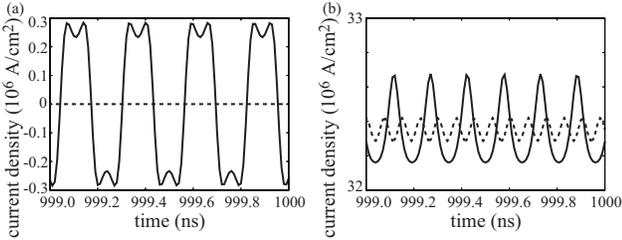}}
\caption{
        (a) The transverse electric current densities given by Eq. (\ref{eq:electric_current_y_total}) 
            for in-phase synchronization (solid line) and antiphase synchronization (dotted line). 
        (b) The longitudinal electric current densities given by Eq. (\ref{eq:electric_current_x_total}) 
            for in-phase or antiphase synchronization (solid line) and phase difference of a quarter of a period (dotted line). 
         \vspace{-3ex}}
\label{fig:fig6}
\end{figure}




The phase difference can be measured from the spin Hall magnetoresistance effect. 
According to Eq. (\ref{eq:electric_current_y}), the total electric current density in the transverse direction is 
\begin{equation}
\begin{split}
  J_{\rm c}^{\rm T}
  &=
  j_{1y}
  +
  j_{3y}
\\
  &=
  -\left(
    \chi
    m_{1x}
    m_{1y}
    +
    \chi^{\prime}
    m_{1z}
  \right)
  J_{0}
\\
  &\ \ -
  \left(
    \chi
    m_{3x}
    m_{3y}
    +
    \chi^{\prime}
    m_{3z}
  \right)
  J_{0}. 
  \label{eq:electric_current_y_total}
\end{split}
\end{equation}
Then, an oscillating current appears in the transverse direction for in-phase synchronization, 
whereas the transverse current becomes zero for antiphase synchronization, as shown in Fig. \ref{fig:fig6}(a). 
The Fourier transformation of Eq. (\ref{eq:electric_current_y_total}) for in-phase synchronization has peaks at 
the frequencies of $f_{n}=(2n-1)f_{0}$ ($n=1,2,3,\cdots$), where $f_{0}$ is the lowest frequency; 
see Appendix \ref{sec:Analytical solution of magnetization on a constant energy curve}. 
Similarly, the total electric current density in the longitudinal direction is 
\begin{equation}
\begin{split}
  J_{\rm c}^{\rm L}
  &=
  J_{0}
  +
  j_{1x}
  +
  j_{2x}
\\
  &=
  J_{0}
  +
  \left(
    \chi^{\prime\prime}
    +
    \chi
    m_{1y}^{2}
  \right)
  J_{0}
  +
  \left(
    \chi^{\prime\prime}
    +
    \chi
    m_{2y}^{2}
  \right)
  J_{0}.
  \label{eq:electric_current_x_total}
\end{split}
\end{equation}
The oscillation frequency of this current is different for the synchronizations 
having the phase difference of in-phase or antiphase and that of a quarter of a period, 
as shown in Fig. \ref{fig:fig6}(b). 
The Fourier transformation of Eq.(\ref{eq:electric_current_x_total}) has the peaks at 
$f_{n}=2n f_{0}$ for in-phase or antiphase and $f_{n}=4n f_{0}$ when the phase difference is a quarter of a period. 


An interesting question regarding these numerical results is to clarify the reason why 
the phase difference finally becomes a certain value for a given set of the parameters, i.e., 
which phase difference is an attractor of the limit cycle. 
It is, however, difficult to answer this question directly due to the following reason. 
We note that the present model includes several small-valued parameters, $\alpha$, $\beta$, $\chi$, $\chi^{\prime}$, and $\chi^{\prime\prime}$, 
as shown in Eqs. (\ref{eq:LLG_L_exact}) and (\ref{eq:LLG_T_exact}), and is complicated. 
The torque related to the lowest order of $\beta$ in these equations is the conventional field-like torque 
given by $[\gamma\beta \hbar \vartheta_{\rm R} J_{0}/(2eMd_{\rm F})] \mathbf{m}_{\ell} \times \mathbf{e}_{y}$. 
It should be noted that the phase difference is not determined solely by this term because this lowest order term of the field-like torque 
does not include the coupling between the magnetizations. 
Similarly, the attractor is not determined solely by the lowest order term of the coupling torque, which is 
$\left[ \gamma \hbar \vartheta_{\rm R} J_{0}/(2eMd_{\rm F}) \right] \chi m_{\ell^{\prime} y}^{2} \mathbf{m}_{\ell} \times \left( \mathbf{e}_{y} \times \mathbf{m}_{\ell} \right)$ in Eq. (\ref{eq:LLG_L_exact}) and 
$\left[ \gamma \hbar \vartheta_{\rm R} J_{0}/(2eMd_{\rm F}) \right] \chi m_{\ell^{\prime} x} m_{\ell^{\prime} y} \mathbf{m}_{\ell} \times \left( \mathbf{e}_{x} \times \mathbf{m}_{\ell} \right)$ in Eq. (\ref{eq:LLG_T_exact}), 
because this torque does not include the field-like torque. 
Their combinations or the higher order terms including both $\beta$ and the coupling torques 
related to $\chi$, $\chi^{\prime}$, and $\chi^{\prime\prime}$ should be taken into account 
to answer the question, which is difficult due to the nonlinearity and complexity of the LLG equation. 


Nevertheless, it is possible to reveal the relation between the current and frequency in the synchronized oscillation state 
by assuming a certain value of the phase difference between the magnetizations. 
The current-frequency relation has been often measured in the experiments, 
and therefore, it will be useful to develop a theory clarifying the role of the coupling on the current-frequency relation. 
In the next section, we will investigate this subject. 







\section{Theoretical analysis of current-frequency relation}
\label{sec:Theoretical analysis of current-frequency relation}

The purpose of this section is to develop an analytical theory of the synchronization 
revealing the relation among the current, frequency, and the phase difference of the magnetizations 
in the synchronized oscillation state. 


\subsection{Basis of analysis}

Here, let us mention the basis of our theoretical analysis. 
It is difficult to solve the LLG equation exactly because of its nonlinearity.  
Instead, we employ the averaging technique of the LLG equation on constant energy curves \cite{bertotti09}. 
This approach has been used to study 
the spin torque switching in thermally activated regions \cite{apalkov05,bertotti06,newhall13,taniguchi13,pinna13} 
and spin torque oscillators \cite{bertotti05,serpico05,ebels08,bazaliy11,dykman12,lacoste13,taniguchi14APEX,pinna14,taniguchi15PRB}, 
as well as the microwave assisted magnetization reversal \cite{taniguchi14,suto15}, 
but has not been applied to the coupled system. 
This approach is valid when the magnetic energy is changed slowly compared with the oscillation period. 
We note that only the lowest order terms in the LLG equation is necessary to derive the current-frequency relation, 
as far as several parameters such as $\beta$ are small. 
Thus, we use the simplified LLG equation maintaining only the dominant terms. 
The LLG equation used in this section for the longitudinal geometry is 
\begin{equation}
\begin{split}
  \frac{d \mathbf{m}_{\ell}}{dt}
  \simeq
  &
  -\gamma
  \mathbf{m}_{\ell}
  \times
  \mathbf{H}_{\ell}
  -
  \alpha
  \gamma
  \mathbf{m}_{\ell}
  \times
  \left(
    \mathbf{m}_{\ell}
    \times
    \mathbf{H}_{\ell}
  \right)
\\
  &-
  \frac{\gamma \hbar \vartheta_{\rm R} J_{0}}{2eMd_{\rm F}}
  \left(
    1
    +
    \chi^{\prime\prime}
    +
    \chi
    m_{\ell^{\prime} y}^{2}
  \right)
  \mathbf{m}_{\ell}
  \times
  \left(
    \mathbf{e}_{y}
    \times
    \mathbf{m}_{\ell}
  \right),
  \label{eq:LLG_L_approx}
\end{split}
\end{equation}
whereas that for the transverse geometry is 
\begin{equation}
\begin{split}
  \frac{d \mathbf{m}_{\ell}}{dt}
  \simeq&
  -\gamma
  \mathbf{m}_{\ell}
  \times
  \mathbf{H}_{\ell}
  -
  \alpha
  \gamma
  \mathbf{m}_{\ell}
  \times
  \left(
    \mathbf{m}_{\ell}
    \times
    \mathbf{H}_{\ell}
  \right)
\\
  &-
  \frac{\gamma \hbar \vartheta_{\rm R} J_{0}}{2eMd_{\rm F}}
  \mathbf{m}_{\ell}
  \times
  \left(
    \mathbf{e}_{y}
    \times
    \mathbf{m}_{\ell}
  \right)
\\
  &-
  \frac{\gamma \hbar \vartheta_{\rm R} J_{0}}{2eMd_{\rm F}}
  \chi 
  m_{\ell^{\prime} x}
  m_{\ell^{\prime} y}
  \mathbf{m}_{\ell}
  \times
  \left(
    \mathbf{e}_{x}
    \times
    \mathbf{m}_{\ell}
  \right).
  \label{eq:LLG_T_approx}
\end{split}
\end{equation}


In the self-oscillation state, the damping torque during the precession is balanced with the spin torque, 
and the torque due to the magnetic field, corresponding to the first term on the right hand side of Eq. (\ref{eq:LLG}), 
becomes the dominant term determining the magnetization dynamics. 
The dynamic trajectory given by this field torque corresponds to constant energy curves of the energy density, 
$E=-M \int d \mathbf{m}_{\ell}\cdot\mathbf{H}_{\ell}$, where its explicit form is 
\begin{equation}
  E
  =
  -\frac{MH_{\rm K}}{2}
  m_{\ell y}^{2}
  +
  2\pi M^{2}
  m_{\ell z}^{2}. 
  \label{eq:energy}
\end{equation}
The minimum and saddle points of Eq. (\ref{eq:energy}) are 
$\mathbf{m}_{\rm min}=\pm \mathbf{e}_{y}$ and $\mathbf{m}_{\rm saddle}=\pm \mathbf{e}_{x}$, 
where the corresponding energy densities are $E_{\rm min}=-MH_{\rm K}/2$ and $E_{\rm saddle}=0$. 
The solution of $\mathbf{m}_{\ell}$ precessing on a constant energy curve is described by the Jacobi elliptic function \cite{byrd71} as 
(see also Appendix \ref{sec:Analytical solution of magnetization on a constant energy curve} summarizing the derivations) 
\begin{equation}
  m_{x}
  =
  \sqrt{
    1
    +
    \frac{2E}{MH_{\rm K}}
  }
  {\rm sn}
  \left[
    \frac{4 \mathsf{K}(k)}{\tau(E)} t
    +
    \varphi_{0},
    k
  \right],
  \label{eq:mx_sol}
\end{equation}
\begin{equation}
  m_{y}
  =
  \sqrt{
    \frac{4\pi M-2E/M}{H_{\rm K}+4\pi M}
  }
  {\rm dn}
  \left[
    \frac{4 \mathsf{K}(k)}{\tau(E)} t
    +
    \varphi_{0},
    k
  \right],
  \label{eq:my_sol}
\end{equation}
\begin{equation}
  m_{z}
  =
  \sqrt{
    \frac{H_{\rm K}+2E/M}{H_{\rm K}+4\pi M}
  }
  {\rm cn}
  \left[
    \frac{4 \mathsf{K}(k)}{\tau(E)} t
    +
    \varphi_{0},
    k
  \right],
  \label{eq:mz_sol}
\end{equation}
where $\varphi_{0}$ is the initial phase. 
The period of $m_{x}$ and $m_{z}$ is $\tau(E)$, whereas that of $m_{y}$ is $\tau(E)/2$, 
where the precession frequency $f(E)=1/\tau(E)$ is given by 
\begin{equation}
  f(E) 
  =
  \frac{\gamma \sqrt{H_{\rm K} \left( 4 \pi M - 2E/M \right)}}{4 \mathsf{K}(k)}, 
  \label{eq:frequency}
\end{equation}
where $\mathsf{K}(k)$ is the first kind of complete elliptic integral 
with the modulus 
\begin{equation}
  k
  =
  \sqrt{
    \frac{4 \pi M \left( H_{\rm K} + 2E/M \right)}{H_{\rm K} \left( 4 \pi M - 2E/M \right)}
  }.
  \label{eq:modulus}
\end{equation}
Note that Eq. (\ref{eq:frequency}) reproduces the ferromagnetic resonance (FMR) frequency, 
$\gamma\sqrt{H_{\rm K} \left( H_{\rm K} + 4\pi M \right)}/(2\pi)$, 
in the limit of $E \to E_{\rm min}$. 
Identifying $E$ and $\varphi_{0}$ corresponds to the determination of the initial condition. 


The averaged technique investigates the energy change during a precession on a constant energy curve, 
which is obtained from the LLG equation as 
\begin{equation}
  \oint 
  dt 
  \frac{dE}{dt}
  =
  \mathscr{W}_{\rm s}
  +
  \mathscr{W}_{\rm s}^{\rm L,T}
  +
  \mathscr{W}_{\alpha},
  \label{eq:LLG_averaged}
\end{equation}
where $\mathscr{W}_{\rm s}$ is the energy change by the conventional spin torque due to the spin Hall effect, 
whereas $\mathscr{W}_{\alpha}$ is the dissipation due to the damping torque. 
The integral range is over the precession period. 
The explicit forms of $\mathscr{W}_{\rm s}$ and $\mathscr{W}_{\alpha}$ are given by \cite{taniguchi13}
\begin{equation}
\begin{split}
  \mathscr{W}_{\rm s}
  &=
  \oint 
  dt 
  \frac{\gamma \hbar \vartheta_{\rm R} J_{0}}{2ed_{\rm F}}
  \left[
    \mathbf{e}_{y}
    \cdot
    \mathbf{H}_{\ell}
    -
    \left(
      \mathbf{m}_{\ell}
      \cdot
      \mathbf{e}_{y}
    \right)
    \left(
      \mathbf{m}_{\ell}
      \cdot
      \mathbf{H}_{\ell}
    \right)
  \right]
\\
  &=
  \frac{\pi \hbar \theta_{\rm R} J_{0} \left( H_{\rm K} + 2E/M \right)}{ed_{\rm F} \sqrt{H_{\rm K} \left( H_{\rm K} + 4 \pi M \right)}},
  \label{eq:Melnikov_s}
\end{split}
\end{equation}
\begin{equation}
\begin{split}
  \mathscr{W}_{\alpha}
  &=
  -\oint 
  dt 
  \alpha 
  \gamma 
  M 
  \left[
    \mathbf{H}_{\ell}^{2}
    -
    \left(
      \mathbf{m}_{\ell}
      \cdot
      \mathbf{H}_{\ell}
    \right)^{2}
  \right]
\\
  &=
  -4 \alpha M 
  \sqrt{\frac{4\pi M-2E/M}{H_{\rm K}}}
  \left[
    \frac{2E}{M}
    \mathsf{K}(k)
    +
    H_{\rm K}
    \mathsf{E}(k)
  \right],
  \label{eq:Melnikov_alpha}
\end{split}
\end{equation}
where $\mathsf{E}(k)$ is the second kind of complete elliptic integral. 


On the other hand, $\mathscr{W}_{\rm s}^{\rm L,T}$ represents the work done by the coupling torque 
in the longitudinal or transverse geometry, corresponding to the last term in Eq. (\ref{eq:LLG_L_approx}) or (\ref{eq:LLG_T_approx}). 
The explicit forms of $\mathscr{W}_{\rm s}^{\rm L}$ and $\mathscr{W}_{\rm s}^{\rm T}$ are shown in the following sections. 
For both cases, the relation between the current and frequency is clarified as follows. 
In the self-oscillation state, since the spin torque balances the damping torque, the following condition should be satisfied, 
\begin{equation}
  \oint 
  dt 
  \frac{dE}{dt}
  =
  0.
  \label{eq:oscillation_condition}
\end{equation} 
The current density $J_{0}$ satisfying Eq. (\ref{eq:oscillation_condition}) is the current necessary to excite the self-oscillation 
on the constant energy curve of $E$, and is denoted as $J_{0}(E)$. 
The relation between the current and frequency in the self-oscillation state is given by 
this $J_{0}(E)$ and $f(E)$ given by Eq. (\ref{eq:frequency}). 
It should be emphasized that the current density $J_{0}(E)$ depends on the phase difference between the magnetizations 
through $\mathscr{W}_{\rm s}^{\rm L,T}$. 
We will therefore study the relation between the phase difference and the current-frequency relation in line with this deduction. 


\subsection{Transverse geometry}

In this section, we investigate the current-frequency relation in the transverse geometry. 
The work done by the coupling torque is defined as 
\begin{equation}
  \mathscr{W}_{\rm s}^{\rm T}
  =\!
  \oint 
  dt 
  \frac{\gamma \hbar \vartheta_{\rm R} \chi J_{0}}{2ed_{\rm F}}
  m_{\ell^{\prime} x}
  m_{\ell^{\prime} y}
  \!\!
  \left[
    \mathbf{e}_{x}
    \cdot
    \mathbf{H}_{\ell}
    -
    \left(
      \mathbf{m}_{\ell}
      \cdot
      \mathbf{e}_{x}
    \right)
    \left(
      \mathbf{m}_{\ell}
      \cdot
      \mathbf{H}_{\ell}
    \right)
  \right].
  \label{eq:Melnikov_T_def} 
\end{equation}


Before advancing the calculation, let us briefly mention the definition of the phase difference in the present approach. 
If the oscillation trajectory is a circle, the phase difference is easily defined, 
i.e., the antiphase corresponds to $\Delta\varphi=\pi$, whereas $\Delta\varphi=0$ is the in-phase. 
In the present case, on the other hand, the oscillation trajectory is described by the elliptic function, 
as shown in Eqs. (\ref{eq:mx_sol}), (\ref{eq:my_sol}), and (\ref{eq:mz_sol}). 
In this case, the phase difference is defined using the elliptic integral $\mathsf{K}(k)$, 
where $\Delta\varphi=0$ for the in-phase synchronization, and the antiphase synchronization corresponds to $\Delta\varphi=2 \mathsf{K}(k)$. 
It is useful to note that $\Delta\varphi=2 \mathsf{K}(k)$ becomes $\pi$ in the limit of $k \to 0$, 
corresponding to the case that the oscillation trajectory becomes a circle. 
Similarly, $\Delta\varphi=\mathsf{K}(k)$ means that the phase difference is a quarter of a period. 


Equation (\ref{eq:Melnikov_T_def}) for an arbitrary phase difference is evaluated 
by numerically calculating the integral with the solutions of $\mathbf{m}_{\ell}$ and $\mathbf{m}_{\ell^{\prime}}$ 
shown in Appendix \ref{sec:Details of calculations}. 
It is, however, useful to derive the analytical solutions of Eq. (\ref{eq:Melnikov_T_def}) 
for specific values of the phase difference. 
Equation (\ref{eq:Melnikov_T_def}) for both the in-phase ($\Delta\varphi=0$) and antiphase [$\Delta\varphi=2 \mathsf{K}(k)$] becomes 
\begin{equation}
  \mathscr{W}_{\rm s}^{\rm T}
  =
  \mp
  \frac{\pi \hbar \vartheta_{\rm R} \chi J_{0}}{2 ed_{\rm F} \sqrt{H_{\rm K} \left( H_{\rm K} + 4 \pi M \right)}}
  \left(
    -\frac{2E}{M}
  \right)
  \left(
    1
    +
    \frac{2E}{MH_{\rm K}}
  \right),
  \label{eq:Melnikov_T_sol}
\end{equation}
where the double sign means the upper for the in-phase ($\Delta\varphi=0$) synchronization 
and the lower for the antiphase [$\Delta\varphi=2 \mathsf{K}(k)$] synchronization. 
Equation (\ref{eq:Melnikov_T_sol}) is zero at $E=E_{\rm min}$ and $E_{\rm saddle}$, 
and is negative (positive) for the energy density $E$ in the rage of $E_{\rm min}<E<E_{\rm saddle}$ 
when $\Delta\varphi=0$ [$2 \mathsf{K}(k)$]. 
This means that the coupling torque acts as a damping (an antidamping) torque 
when the phase difference is in-phase (antiphase). 
We also note that $\mathscr{W}_{\rm s}^{\rm T}=0$ when 
$\Delta\varphi= \mathsf{K}(k)$; i.e., the phase difference is a quarter of a period. 
The calculations necessary to obtain these specific values of $\mathscr{W}_{\rm s}^{\rm T}$ are 
also summarized in Appendix \ref{sec:Details of calculations}. 
We note that the sign change of $\mathscr{W}_{\rm s}^{\rm T}$ with respect to the phase difference 
is related to the fact that the coupling torque in the transverse geometry, Eq. (\ref{eq:torque_T}), 
has the angular dependence of $m_{\ell^{\prime} x}m_{\ell^{\prime} y} \mathbf{m}_{\ell} \times \left( \mathbf{e}_{x} \times \mathbf{m}_{\ell} \right)$. 
Because of this angular dependence, the coupling torque acts as an anti-damping (a damping) torque 
when $m_{\ell x}$ and $m_{\ell^{\prime} x}$ have the opposite (same) signs, 
resulting in the increase (decrease) of the energy supplied to the ferromagnets by the coupling torque. 


In summary, the work done by the coupling torque, $\mathscr{W}_{\rm s}^{\rm T}$, 
is negative and minimized at the in-phase ($\Delta\varphi=0$), 
zero for $\Delta\varphi=\mathsf{K}(k)$, 
and positive and maximized at the antiphase [$\Delta\varphi=2\mathsf{K}(k)$].




\begin{figure}
\centerline{\includegraphics[width=1.0\columnwidth]{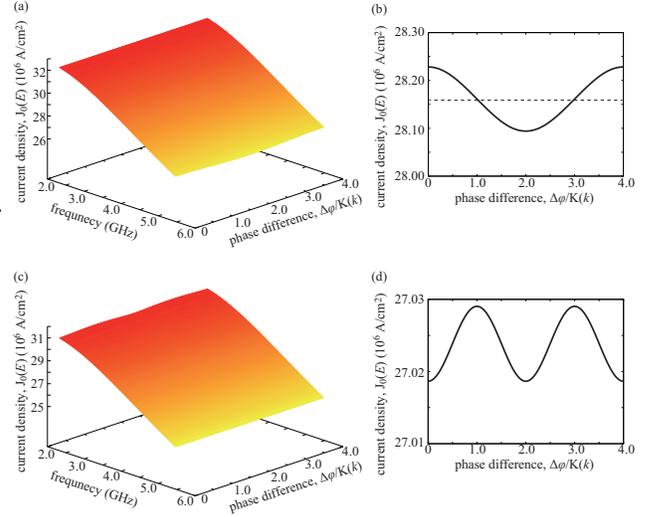}}
\caption{
         (a) The current density, $J_{0}(E)$, necessary to excite the self-oscillation in the transverse geometry 
             as a function of the oscillation frequency $f(E)$ and the phase difference $\Delta\varphi$ of the magnetizations. 
             The phase difference is in the unit of $\mathsf{K}(k)$. 
         (b) Dependence of $J_{0}(E)$ for the transverse geometry, on $\Delta\varphi$ at $f(E)=4.6$ GHz. 
             The dotted line represents $J_{0}(E)$ in the absence of the coupling. 
         (c) The relation among $J_{0}(E)$, $f(E)$, and $\Delta\varphi$ in the longitudinal geometry. 
         (d) The current density $J_{0}(E)$ for $f(E)=4.6$ GHz in the longitudinal geometry. 
         \vspace{-3ex}}
\label{fig:fig7}
\end{figure}





The current $J_{0}$ in the transverse geometry is defined as 
\begin{equation}
  J_{0}(E)
  =
  \frac{2 \alpha eMd_{\rm F}}{\hbar \vartheta_{\rm R}}
  \frac{\mathcal{N}}{\mathcal{D}_{\rm T}},
  \label{eq:jE_def}
\end{equation}
where $\mathcal{N}$ and $\mathcal{D}_{\rm T}$ are defined as 
\begin{equation}
  \mathcal{N}
  =
  \gamma
  \oint 
  dt 
  \left[
    \mathbf{H}_{\ell}^{2}
    -
    \left(
      \mathbf{m}_{\ell}
      \cdot
      \mathbf{H}_{\ell}
    \right)^{2}
  \right],
\end{equation}
\begin{equation}
\begin{split}
  \mathcal{D}_{\rm T}
  =&
  \gamma
  \oint 
  dt 
  \left[
    \mathbf{e}_{y}
    \cdot
    \mathbf{H}_{\ell}
    -
    \left(
      \mathbf{m}_{\ell}
      \cdot
      \mathbf{e}_{y}
    \right)
    \left(
      \mathbf{m}_{\ell}
      \cdot
      \mathbf{H}_{\ell}
    \right)
  \right]
\\
  &+
  \gamma 
  \chi 
  \oint 
  dt 
  m_{\ell^{\prime} x}
  m_{\ell^{\prime} y}
  \left[
    \mathbf{e}_{x}
    \cdot
    \mathbf{H}_{\ell}
    -
    \left(
      \mathbf{m}_{\ell}
      \cdot
      \mathbf{e}_{x}
    \right)
    \left(
      \mathbf{m}_{\ell}
      \cdot
      \mathbf{H}_{\ell}
    \right)
  \right]. 
\end{split}
\end{equation}
The explicit form of $\mathcal{N}$ is obtained from Eq. (\ref{eq:Melnikov_alpha}) as 
\begin{equation}
  \mathcal{N}
  =
  4 
  \sqrt{
    \frac{4\pi M-2E/M}{H_{\rm K}}
  }
  \left[
    \frac{2E}{M}
    \mathsf{K}(k)
    +
    H_{\rm K}
    \mathsf{E}(k)
  \right].
\end{equation}
On the other hand, 
$\mathcal{D}_{\rm T}$ for the in-phase or antiphase is obtained from Eqs. (\ref{eq:Melnikov_s}) and (\ref{eq:Melnikov_T_sol}) as 
\begin{equation}
\begin{split}
  \mathcal{D}_{\rm T}
  =&
  \frac{2\pi \left( H_{\rm K} + 2E/M \right)}{\sqrt{H_{\rm K} \left( H_{\rm K} + 4 \pi M \right)}}
\\
  &
  \mp
  \frac{\pi \chi \left(-2E/M \right) \left[ 1 + 2E/\left( MH_{\rm K} \right) \right]}{\sqrt{H_{\rm K} \left( H_{\rm K} + 4 \pi M \right)}},
\end{split}
\end{equation}
where the double sign means the upper for the in-phase synchronization 
and the lower for the antiphase synchronization. 


Figure \ref{fig:fig7}(a) shows $J_{0}(E)$ as functions of the oscillation frequency $f(E)$ and the phase difference $\Delta\varphi$. 
The current-frequency relation in the transverse geometry can be obtained from this figure. 
To reveal the role of the phase difference more clearly, 
we show $J_{0}(E)$ for a certain value of $f(E)(=4.6 {\rm GHz})$ in Fig. \ref{fig:fig7}(b). 
Note that $J_{0}(E)$ is smaller than that in the absence of the coupling, which is shown by the dotted line, 
and minimized when $\Delta\varphi=2 \mathsf{K}(k)$, i.e., the antiphase. 
This is because the work done by the coupling torque is positive and maximized at the antiphase. 
On the other hand, $J_{0}(E)$ is maximized at the in-phase, and is larger than that in the absence of the coupling 
because the work done by the coupling torque is negative and minimized at the in-phase. 
We notice that the phase differences observed in the numerical simulation in Sec. \ref{sec:Numerical analysis of synchronization}, 
i.e., the in-phase and antiphase, 
correspond to $\Delta\varphi$ satisfying 
\begin{equation}
  \frac{\partial J_{0}}{\partial \Delta\varphi}
  =
  0, 
\end{equation}
or equivalently, 
\begin{equation}
  \frac{\partial \mathscr{W}_{\rm s}^{\rm T}}{\partial \Delta\varphi}
  =
  0. 
  \label{eq:stable_phase_condition_T}
\end{equation}
In other words, the phase differences observed in the numerical simulations correspond to those 
giving the extrema of $J_{0}$ ($\mathscr{W}_{\rm s}^{\rm T}$).



\subsection{Longitudinal geometry}

Let us investigate the theoretical relation between the current and frequency in the longitudinal geometry. 
In this case, the averaged LLG equation is given by 
\begin{equation}
  \oint
  dt 
  \frac{dE}{dt}
  =
  \left(
    1
    +
    \chi^{\prime\prime}
  \right)
  \mathscr{W}_{\rm s}
  +
  \mathscr{W}_{\rm s}^{\rm L}
  +
  \mathscr{W}_{\alpha}, 
\end{equation}
where $\mathscr{W}_{\rm s}$ and $\mathscr{W}_{\alpha}$ are given by Eqs. (\ref{eq:Melnikov_s}) and (\ref{eq:Melnikov_alpha}). 
On the other hand, $\mathscr{W}_{\rm s}^{\rm L}$ representing the energy change due to the longitudinal coupling is defined as 
\begin{equation}
  \mathscr{W}_{\rm s}^{\rm L}
  =\!
  \oint 
  dt 
  \frac{\gamma \hbar \vartheta_{\rm R} \chi J_{0}}{2ed_{\rm F}}
  m_{\ell^{\prime} y}^{2}
  \!\!
  \left[
    \mathbf{e}_{y}
    \cdot
    \mathbf{H}_{\ell}
    -
    \left(
      \mathbf{m}_{\ell}
      \cdot
      \mathbf{e}_{y}
    \right)
    \left(
      \mathbf{m}_{\ell}
      \cdot
      \mathbf{H}_{\ell}
    \right)
  \right].
  \label{eq:Melnikov_L_def} 
\end{equation}
For both the in-phase and antiphase, Eq. (\ref{eq:Melnikov_L_def}) becomes 
(see Appendix \ref{sec:Details of calculations})
\begin{equation}
\begin{split}
  \mathscr{W}_{\rm s}^{\rm L}
  =
  &
  \frac{\pi \hbar \vartheta_{\rm R} \chi J_{0}}{2ed_{\rm F} \sqrt{H_{\rm K} \left( H_{\rm K} + 4 \pi M \right)}}
\\
  & \times
  \frac{\left( H_{\rm K} + 2 E/M \right) \left[ 4 \pi M \left( H_{\rm K} - 2E/M \right) - 2 H_{\rm K} \left(2E/M \right) \right]}{H_{\rm K} \left( H_{\rm K} + 4 \pi M \right)}. 
  \label{eq:Melnikov_L_sol}
\end{split}
\end{equation}
On the other hand, when the phase difference is a quarter of a period [$\Delta\varphi=\mathsf{K}(k)$], Eq. (\ref{eq:Melnikov_L_def}) becomes 
\begin{equation}
  \mathscr{W}_{\rm s}^{\rm L}
  =
  \frac{\pi \hbar \vartheta_{\rm R} \chi J_{0}}{ed_{\rm F}}
  \frac{H_{\rm K}+2E/M}{H_{\rm K} \left( H_{\rm K} + 4 \pi M \right)}
  \sqrt{
    -\frac{2E}{M}
    \left(
      4\pi M 
      -
      \frac{2E}{M}
    \right)
  }.
  \label{eq:Melnikov_L_sol_quarter}
\end{equation}
It should be emphasized that $\mathscr{W}_{\rm s}^{\rm L}$ is always positive for an arbitrary phase difference. 
This is because the coupling torque in Eq. (\ref{eq:torque_L}) always acts as an anti-damping torque. 


We can determine the current density $J_{0}(E)$ satisfying Eq. (\ref{eq:oscillation_condition}) in the longitudinal geometry, 
as in the case of the transverse geometry, 
by replacing $\mathcal{D}_{\rm T}$ in Eq. (\ref{eq:jE_def}) with 
\begin{equation}
\begin{split}
  \mathcal{D}_{\rm L}
  =&
  \gamma
  \left(
    1 
    + 
    \chi^{\prime\prime}
  \right) 
  \oint 
  dt 
  \left[
    \mathbf{e}_{y}
    \cdot
    \mathbf{H}_{\ell}
    -
    \left(
      \mathbf{m}_{\ell}
      \cdot
      \mathbf{e}_{y}
    \right)
    \left(
      \mathbf{m}_{\ell}
      \cdot
      \mathbf{H}_{\ell}
    \right)
  \right]
\\
  &+
  \gamma 
  \chi 
  \oint 
  dt 
  m_{\ell^{\prime} y}^{2}
  \left[
    \mathbf{e}_{y}
    \cdot
    \mathbf{H}_{\ell}
    -
    \left(
      \mathbf{m}_{\ell}
      \cdot
      \mathbf{e}_{y}
    \right)
    \left(
      \mathbf{m}_{\ell}
      \cdot
      \mathbf{H}_{\ell}
    \right)
  \right]. 
\end{split}
\end{equation}
The explicit form of $\mathcal{D}_{\rm L}$ for both the in-phase and antiphase is obtained from Eqs. (\ref{eq:Melnikov_s}) and (\ref{eq:Melnikov_L_sol}) as 
\begin{equation}
\begin{split}
  \mathcal{D}_{\rm L}
  &=
  \frac{2\pi \left( 1 + \chi^{\prime\prime} \right) \left( H_{\rm K} + 2E/M \right)}{\sqrt{H_{\rm K} \left( H_{\rm K} + 4 \pi M \right)}}
\\
  &+
  \frac{\pi \chi \left( H_{\rm K} + 2E/M \right) \left[ 4 \pi M \left( H_{\rm K} - 2E/M \right) - 2 H_{\rm K} \left( 2E/M \right) \right]}{ \left[ H_{\rm K} \left( H_{\rm K} + 4\pi M \right) \right]^{3/2}},
\end{split}
\end{equation}
whereas that when the phase difference is a quarter of a period is obtained from Eqs. (\ref{eq:Melnikov_s}) and (\ref{eq:Melnikov_L_sol_quarter}) as 
\begin{equation}
\begin{split}
  \mathcal{D}_{\rm L}
  =&
  \frac{2 \pi \left( 1 + \chi^{\prime\prime} \right) \left( H_{\rm K} + 2E/M \right)}{\sqrt{H_{\rm K} \left( H_{\rm K} + 4\pi M \right)}}
\\
  &+
  \frac{2\pi \chi \left( H_{\rm K} + 2E/M \right)}{H_{\rm K} \left( H_{\rm K} + 4\pi M \right)}
  \sqrt{
    -\frac{2E}{M}
    \left(
      4\pi M 
      -
      \frac{2E}{M}
    \right)
  }.
\end{split}
\end{equation}


Figure \ref{fig:fig7}(c) shows $J_{0}(E)$ as functions of $f(E)$ and $\Delta\varphi$. 
The current-frequency relation in the longitudinal geometry can be obtained from this figure. 
It is noted that $J_{0}(E)$ is always smaller than that in the absence of the coupling 
because the coupling torque in the longitudinal geometry always points to the anti-damping direction, 
and therefore, the work done by the coupling torque is always positive. 
Figure \ref{fig:fig7}(b) shows $J_{0}(E)$ as a function of $\Delta\varphi$ at a certain value of $f(E)$.
As shown, $J_{0}(E)$ has minima at both the in-phase ($\Delta\varphi=0$) and the antiphase [$\Delta\varphi=2\mathsf{K}(k)$], 
whereas it is maximized when the phase difference is a quarter of a period [$\Delta\varphi=\mathsf{K}(k)$]. 
We again notice that these phase differences found in the numerical simulations in Sec. \ref{sec:Numerical analysis of synchronization} 
correspond to $\Delta\varphi$ satisfying $\partial J_{0}(E)/\partial \Delta\varphi=0$, or equivalently, 
\begin{equation}
  \frac{\partial \mathscr{W}_{\rm s}^{\rm L}}{\partial \Delta\varphi}
  =
  0.
  \label{eq:stable_phase_condition_L}
\end{equation}


\subsection{Phase differences in stable synchronization and fixed points of effective potential}

Equation (\ref{eq:LLG_averaged}) describes the slow change of the magnetic energy in the oscillation state. 
The magnetization dynamics is regarded as a motion of a point particle in an effective potential given by its right-hand side. 
Equations (\ref{eq:stable_phase_condition_T}) and (\ref{eq:stable_phase_condition_L}) correspond to the stability conditions of the point particle in this effective potential. 
Therefore, the phase difference found in the numerical simulation finally converges to 
one of these values satisfying Eq. (\ref{eq:stable_phase_condition_T}) or Eq. (\ref{eq:stable_phase_condition_L}). 
Whether the in-phase, antiphase, or the phase difference with a quarter of a period becomes 
the attractor depends on the higher order terms of the small parameters, as well as the initial states of the magnetizations, 
as mentioned at the end of Sec. \ref{sec:Numerical analysis of synchronization}. 
This discussion is beyond the scope of this paper.


\subsection{Instability threshold}

At mentioned at the beginning of Sec. \ref{sec:Numerical analysis of synchronization}, 
the in-plane self-oscillation for a single ferromagnet is stabilized 
when the current density is in the range of $J_{\rm c} < J_{0} < J^{*}$, 
where $J_{\rm c}$ and $J^{*}$ are given by Eqs. (\ref{eq:J_c_single}) and (\ref{eq:J_switch_single}), respectively. 
At the end of this section, let us briefly discuss the effect of the coupling on these scaling currents. 


Let us remind the readers that $J_{\rm c}$ is the current density necessary to destabilize the magnetization in equilibrium, 
whereas $J^{*}$ is the current necessary to overcome the energy barrier, $E_{\rm saddle}-E_{\rm min}$. 
These current densities are theoretically defined as \cite{taniguchi13}
\begin{equation}
  J_{\rm c}
  =
  \lim_{E \to E_{\rm min}}
  J_{0}(E),
\end{equation}
\begin{equation}
  J^{*}
  =
  \lim_{E \to E_{\rm saddle}}
  J_{0}(E). 
\end{equation}
It is confirmed that Eqs. (\ref{eq:J_c_single}) and (\ref{eq:J_switch_single}) are reproduced 
by substituting Eqs. (\ref{eq:Melnikov_s}) and (\ref{eq:Melnikov_alpha}) in the definition of $J_{0}(E)$ 
in the absence of the coupling. 


On the other hand, in the presence of the transverse coupling, 
it is confirmed from Eq. (\ref{eq:Melnikov_T_sol}) that 
a factor $\left[1 - \left( \chi/2 \right) \right]$ should be multiplied to the denominator of Eq. (\ref{eq:J_c_single}) 
when the phase difference between the magnetizations is in-phase, 
whereas this factor is replaced by $\left[1 + \left( \chi/2 \right) \right]$ when the phase difference is antiphase. 
The other scaling current, $J^{*}$, is unchanged for these phase differences. 
In the longitudinal geometry, we see from Eqs. (\ref{eq:Melnikov_L_sol}) and (\ref{eq:Melnikov_L_sol_quarter}) that 
the factor $\left(1+\chi + \chi^{\prime\prime} \right)$ should be multiplied to the denominator of Eq. (\ref{eq:J_c_single}) 
when the phase difference is in-phase, antiphase, or a quarter of a period, 
whereas, for $J^{*}$, the factor becomes $1 + \chi^{\prime\prime} + \left( \chi/2 \right) \left[ 4\pi M/\left( H_{\rm K} + 4 \pi M \right) \right]$ for in-phase and antiphase, 
and $1+\chi^{\prime\prime}$ when the phase difference is a quarter of a period. 








\section{Conclusion}
\label{sec:Conclusion}

In conclusion, the coupled magnetization dynamics in the ferromagnets through the spin Hall magnetoresistance effect was investigated. 
The coupling appears in both the longitudinal and transverse directions of the alignment of the ferromagnets. 
The in-phase or antiphase synchronization of the magnetization oscillation was found in the transverse geometry 
by solving the LLG equation numerically. 
On the other hand, in addition to them, 
the synchronization having the phase difference of a quarter of a period is also found in the longitudinal geometry. 
It was shown that these phase differences depend on the values of 
the damping constant and the field-like torque strength. 
The analytical theory revealing the relation among the current, frequency, and phase difference was also developed. 
It was shown that the phase differences observed in the numerical simulations correspond to 
that giving the fixed points of the energy supplied by the coupling torque.


\section*{Acknowledgement}

The author is grateful to Takehiko Yorozu for his contributions to the analytical calculations, 
and to Hitoshi Kubota, Sumito Tsunegi, Yoichi Shiota, Shingo Tamaru, Tazumi Nagasawa, Kiwamu Kudo, and Yoji Kawamura for valuable discussions. 
The author is also thankful to Satoshi Iba, Aurelie Spiesser, Atsushi Sugihara, Takahide Kubota, Hiroki Maehara, and Ai Emura 
for their support and encouragement. 
This work was supported by JSPS KAKENHI Grant-in-Aid for Young Scientists (B) 16K17486. 



\appendix



\section{Values of parameters in numerical simulations}
\label{sec:Values of parameters in numerical simulations}

The exact values of the parameters used in the simulations, evaluated from the parameters found in the experiment \cite{kim16}, are 
$\vartheta_{\rm R}=0.16680863$, $\beta=-0.00973617$, $\chi=0.009525272$, $\chi^{\prime}=-0.000152995$, and 
$\chi^{\prime\prime}=0.03516089$ for $g_{\rm i}^{\uparrow\downarrow}/S=1.0$ nm${}^{-2}$. 
In the main text, $\beta=-0.01$ and $\beta=+0.01$ correspond to $\beta=-0.00973617$ and $\beta=+0.00973617$, respectively. 
Strictly speaking, the change of the value of $g_{\rm i}^{\uparrow\downarrow}$ affects not only $\beta$ but also 
other quantities such as $\vartheta_{\rm R}$, $\chi$, and $\chi^{\prime}$. 
We, however, change the value of $\beta$ only in the numerical simulation, for simplicity, 
because the results do not change significantly unless $|g_{\rm i}^{\uparrow\downarrow}/g_{\rm r}^{\uparrow\downarrow}| \ll 1$. 
The LLG equation with these parameters is solved by using the fourth-order Runge-Kutta method 
from $t=0$ to $t=1$ $\mu$s with the time step of $\Delta t=20$ fs; 
i.e., the number of the time mesh is $N_{\rm t}=5 \times 10^{7}$. 


The present system has two stable states at $\mathbf{m}_{\ell}=\pm \mathbf{e}_{y}$. 
For convention, we assume that the magnetizations initially stay near one equilibrium, $\mathbf{m}_{\ell} = + \mathbf{e}_{y}$. 
For in-phase synchronizations, such as those shown in Figs. \ref{fig:fig2}(c) and \ref{fig:fig2}(d), 
the $z$ components are also synchronized with in-phase, i.e., $m_{\ell z}(t)=m_{\ell^{\prime} z}(t)$. 
On the other hand, for antiphase synchronization shown in, 
for example in Figs. \ref{fig:fig2}(e) and \ref{fig:fig2}(f), 
the $z$ components are also synchronized with antiphase, $m_{\ell z}(t)=-m_{\ell^{\prime} z}(t)$. 


The algorithm evaluating the phase differences shown in Figs. \ref{fig:fig3} and \ref{fig:fig5} 
from the discrete numerical data is as follows. 
We gathered $N_{\rm i}=2^{16}=65536$ data of $\mathbf{m}_{\ell}(t)$ ($\ell=1,2,3$) 
from $t= \left( N_{\rm t}-N_{\rm i}+1 \right) \Delta t$ to $t=N_{\rm t}\Delta t=1$ $\mu$s. 
Then, the averaged periods $T_{\ell}$ of the oscillation of each magnetization were evaluated 
from the peaks of $\mathbf{m}_{\ell}(t)$ in this time range as 
$T_{\ell}= \left[ \left( t_{\ell,a}-t_{\ell,a-1} \right) + \cdots + \left( t_{\ell,2}-t_{\ell,1} \right) \right]/\left( N_{\ell}-1 \right)= \left( t_{\ell,a}-t_{\ell,1} \right)/\left( N_{\ell}-1 \right)$, 
where $N_{\ell}$ is the number of the peaks in $\mathbf{m}_{\ell}(t)$, 
whereas $t_{\ell,a}$ is the time corresponding to the $a$-th peak. 
Then, the phase difference is evaluated as 
$\Delta\varphi=\sum_{a=1}^{N^{\prime}} |t_{\ell,a}-t_{\ell^{\prime},a}| /\left( N^{\prime} \bar{T} \right)$, 
where $N^{\prime}={\rm min} \left[ N_{\ell},N_{\ell^{\prime}} \right]$ and $\bar{T}=\left( T_{\ell} + T_{\ell^{\prime}} \right)/2$ with 
$(\ell,\ell^{\prime})=(1,3)$ for the transverse geometry, whereas that is $(1,2)$ for the longitudinal geometry. 
For the in-phase synchronization, this $\Delta\varphi$ is zero because $t_{\ell,a}=t_{\ell^{\prime},a}$. 
When the phase difference is antiphase, $\Delta\varphi=0.50$ because 
$|t_{\ell,a}-t_{\ell^{\prime},a}|=\bar{T}/2$ in this case. 
Similarly, $\Delta\varphi$ is $0.25$ when the phase difference is a quarter of a period. 


Note that the critical current density to excite the self-oscillation, given by Eq. (\ref{eq:J_c_single}), 
is proportional to the damping constant $\alpha$. 
Therefore, the value of the current density should be increased to observe the self-oscillation 
when $\alpha$ is varied, as in the case of Figs. \ref{fig:fig3} and \ref{fig:fig5}. 
In these figures, $J_{0}$ is assumed as $n \times 30$ MA/cm${}^{2}$ for $\alpha=n \times 0.005$ ($n=1-6$). 


The numerical simulation in Fig. \ref{fig:fig2}(e) indicates that the antiphase synchronization is an attractor for $\beta=+0.01$. 
An exception is that if the initial conditions are set to be identical, 
the final state becomes in-phase synchronization due to the symmetry of the LLG equation with respect to the change of $(\ell,\ell^{\prime}) \to (\ell^{\prime},\ell)$. 
Since Eq. (\ref{eq:stable_phase_condition_T}) is satisfied, 
the phase difference is fixed to in-phase even if it is unstable. 
Similar situations occur in other cases for such specific initial conditions. 



\section{Derivation of coupling torque in transverse geometry}
\label{sec:Derivation of coupling torque in transverse geometry}

In a ferromagnetic/nonmagnetic bilayer, the spin current density flowing in the $i$ direction ($i=x,y,z$) 
with the spin polarization in the $\nu$ direction is related to the electrochemical potential $\bar{\mu}_{\rm N}$ and the spin accumulation $\delta\mu_{\rm N}$ via 
\begin{equation}
  J_{{\rm s}i \nu, {\rm N}}
  =
  -\frac{\hbar \sigma_{\rm N}}{2e^{2}}
  \partial_{i}
  \delta
  \mu_{{\rm N},\nu}
  -
  \frac{\hbar \vartheta \sigma_{\rm N}}{2e^{2}}
  \epsilon_{i \nu j}
  \partial_{j}
  \bar{\mu}_{\rm N}, 
\end{equation}
where $\partial_{j} \bar{\mu}_{\rm N}/e$ is the electric field in the $j$ direction, 
and therefore, $\sigma_{\rm N} \partial_{j} \bar{\mu}_{\rm N}/e$ is the electric current density. 
We assume that this equation is extended to 
\begin{equation}
  J_{{\rm s}z \nu, {\rm N}}^{(\ell)}
  =
  -\frac{\hbar \sigma_{\rm N}}{2e^{2}}
  \partial_{z}
  \delta
  \mu_{{\rm N},\nu}^{(\ell)}
  +
  \frac{\hbar \vartheta}{2e}
  \left(
    J_{0}
    \delta_{\nu y}
    -
    j_{\ell^{\prime} y}
    \delta_{\nu x}
  \right),
  \label{eq:spin_current_T}
\end{equation}
in the transverse geometry, 
where $J_{{\rm s}z \nu,{\rm N}}^{(\ell)}$ is the spin current density flowing near the F${}_{\ell}$/N interface in the $z$ direction 
with the spin polarization in the $\nu$ direction. 
The spin accumulation obeys the diffusion equation, 
and the boundary conditions of the diffusion equation are given by the spin current density at the boundaries. 
Using Eq. (\ref{eq:spin_current_T}), 
the solution of the spin accumulation is given by \cite{taniguchi16PRB} 
\begin{equation}
\begin{split}
  &
  \delta
  \mu_{{\rm N},\nu}^{(\ell)}
  =
  \frac{2\pi}{\left( g_{\rm N}/S \right) \sinh \left( d_{\rm N}/\lambda_{\rm N} \right)}
  \left\{
    -J_{{\rm s}z\nu}^{{\rm F}_{\ell}/{\rm N}}
    \cosh
    \left(
      \frac{z+d_{\rm N}}{\lambda_{\rm N}}
    \right)
  \right.
\\
  &-
  \left.
    \frac{\hbar \vartheta}{2e}
    \left(
      J_{0}
      \delta_{\nu y}
      -
      j_{\ell^{\prime}y}
      \delta_{\nu x}
    \right)
    \left[
      \cosh
      \left(
        \frac{z}{\lambda_{\rm N}}
      \right)
      -
      \cosh
      \left(
        \frac{z+d_{\rm N}}{\lambda_{\rm N}}
      \right)
    \right]
  \right\},
  \label{eq:spin_accumulation_sol_T}
\end{split}
\end{equation}
where we assume that the nonmagnet is in the region of $-d_{\rm N} \le z \le 0$. 
The spin current density, $J_{{\rm s}z\nu}^{{\rm F}_{\ell}/{\rm N}}$, at the F${}_{\ell}$/N interface is given by \cite{chen13,taniguchi16PRB} 
\begin{equation}
\begin{split}
  \mathbf{J}_{\rm s}^{{\rm F}_{\ell}/{\rm N}}
  =&
  \frac{\hbar \vartheta g^{*}}{2e g_{\rm N}}
  \tanh
  \left(
    \frac{d_{\rm N}}{2 \lambda_{\rm N}}
  \right)
  \left(
    J_{0}
    m_{\ell y}
    -
    j_{\ell^{\prime} y}
    m_{\ell x}
  \right)
  \mathbf{m}_{\ell}
\\
  &+
  \frac{\hbar}{2e}
  J_{0}
  \left[
    \vartheta_{\rm R}
    \mathbf{m}_{\ell}
    \times
    \left(
      \mathbf{e}_{y}
      \times
      \mathbf{m}_{\ell}
    \right)
    +
    \vartheta_{\rm I}
    \mathbf{e}_{y}
    \times
    \mathbf{m}_{\ell}
  \right]
\\
  &-
  \frac{\hbar}{2e}
  j_{\ell^{\prime} y}
  \left[
    \vartheta_{\rm R}
    \mathbf{m}_{\ell}
    \times
    \left(
      \mathbf{e}_{x}
      \times
      \mathbf{m}_{\ell}
    \right)
    +
    \vartheta_{\rm I}
    \mathbf{e}_{x}
    \times
    \mathbf{m}_{\ell}
  \right], 
  \label{eq:spin_current_FN_T}
\end{split}
\end{equation}
where the vector notation in boldface represents the direction of the spin polarization, 
whereas the spatial direction of Eq. (\ref{eq:spin_current_FN_T}) is defined as the positive direction, 
i.e., from the nonmagnet to the ferromagnet. 


On the other hand, the electric current density in the nonmagnet flowing in the $i$ direction is given by 
\begin{equation}
  J_{{\rm c}i,{\rm N}}
  =
  \frac{\sigma_{\rm N}}{e}
  \partial_{i}
  \bar{\mu}_{\rm N}
  -
  \frac{\vartheta \sigma_{\rm N}}{e}
  \epsilon_{i j \nu}
  \partial_{j}
  \delta
  \mu_{{\rm N},\nu}.
\end{equation}
In the present case, 
the electric current density near the F${}_{\ell}$/N interface flowing in the $y$ direction becomes 
\begin{equation}
  J_{{\rm c}y,{\rm N}}^{(\ell)}
  =
  j_{\ell^{\prime} y}
  -
  \frac{\vartheta \sigma_{\rm N}}{e}
  \partial_{z}
  \delta
  \mu_{{\rm N},x}^{(\ell)}. 
  \label{eq:electric_current_T}
\end{equation}
Substituting Eqs. (\ref{eq:spin_accumulation_sol_T}) and (\ref{eq:spin_current_FN_T}) into Eq. (\ref{eq:electric_current_T}), 
and averaging along the $z$ direction as $\overline{J_{{\rm c}y,{\rm N}}^{(\ell)}}=(1/d_{\rm N})\int_{-d_{\rm N}}^{0} J_{{\rm c}y,{\rm N}}^{(\ell)} dz$, 
we find that 
\begin{equation}
\begin{split}
  \overline{J_{{\rm c}y,{\rm N}}^{(\ell)}}
  =&
  \left(
    1
    +
    \chi^{\prime\prime}
    +
    \chi
    m_{\ell x}^{2}
  \right)
  j_{\ell^{\prime} y}
\\
  &-
  \left(
    \chi 
    m_{\ell x}
    m_{\ell y}
    +
    \chi^{\prime}
    m_{\ell z}
  \right)
  J_{0}. 
\end{split}
\end{equation}
The conservation law of the electric current along the $y$ direction requires that 
$\overline{J_{{\rm c}y,{\rm N}}^{(\ell)}}=j_{\ell y}+j_{\ell^{\prime} y}$. 
Solving this equation for $\ell=1$ and $3$, we obtain Eq. (\ref{eq:electric_current_y}). 
The spin torque is defined from Eq. (\ref{eq:spin_current_FN_T}) as 
\begin{equation}
  \mathbf{T}_{\ell}
  =
  -\frac{\gamma}{Md_{\rm F}}
  \mathbf{m}_{\ell}
  \times
  \left(
    \mathbf{J}_{\rm s}^{{\rm F}_{\ell}/{\rm N}}
    \times
    \mathbf{m}_{\ell}
  \right). 
  \label{eq:spin_torque_def}
\end{equation}



\section{Analytical solution of magnetization on a constant energy curve}
\label{sec:Analytical solution of magnetization on a constant energy curve}

In this appendix, we shows the derivation of the analytical solution of the magnetization on a constant energy curve. 
For simplicity, we remove the subscript $\ell (=1,2,3)$ distinguishing the ferromagnets. 
The magnetization dynamics on a constant energy curve is described by 
the Landau-Lifshitz (LL) equation 
\begin{equation}
  \frac{d \mathbf{m}}{dt}
  =
  -\gamma
  \mathbf{m}
  \times
  \mathbf{H}.
  \label{eq:LL}
\end{equation}
The magnetic field, Eq. (\ref{eq:field}), is related to the magnetic energy density $E$ via $E=-M \int d\mathbf{m} \cdot \mathbf{H}$, as mentioned in the main text. 
Using the relation $m_{x}^{2}+m_{y}^{2}+m_{z}^{2}=1$, Eq. (\ref{eq:energy}) is rewritten as 
\begin{equation}
  m_{z}^{2}
  +
  \frac{H_{\rm K}}{H_{\rm K}+4\pi M}
  m_{x}^{2}
  =
  \frac{2E/M+H_{\rm K}}{H_{\rm K}+4\pi M}.
\end{equation}
This equation indicates that $m_{z}$ and $m_{x}$ can be expressed as 
$m_{z}=v^{\prime}\cos u$ and $m_{x}=\left( v^{\prime}/v \right) \sin u$, respectively, 
where $v$ and $v^{\prime}$ are defined as 
$v^{2}=H_{\rm K}/\left( H_{\rm K} + 4 \pi M \right)$ and $v^{\prime}=\left( 2E/M+H_{\rm K} \right)/\left( H_{\rm K}+4\pi M \right)$. 
Then, $du/dt=\left( du/d \sin u \right) \left(d \sin u/dt \right)=\left(1/\cos u \right) \left[ d \left( v/v^{\prime} \right) m_{x}/dt \right]=\left( v/m_{z} \right) \left( d m_{x}/dt \right)$, 
which becomes, from Eq. (\ref{eq:LL}), 
\begin{equation}
  \frac{d u}{dt}
  =
  \gamma
  \left(
    H_{\rm K}
    +
    4\pi M 
  \right)
  v m_{y}.
\end{equation}
Introducing new variable $w=\sin u$, this equation gives 
\begin{equation}
  \frac{dw}{\sqrt{\left( 1 - w^{2} \right) \left( 1 - k^{2} w^{2} \right)}}
  =
  \gamma
  \left(
    H_{\rm K}
    +
    4\pi M
  \right)
  v \sqrt{1-v^{\prime 2}}
  dt, 
  \label{eq:elliptic_integral_def}
\end{equation}
where the modulus $k$ is given by Eq. (\ref{eq:modulus}). 
The modulus monotonically varies from $0$ to $1$ 
by changing the energy density $E$ from its minimum $E_{\rm min}$ to saddle $E_{\rm saddle}$. 
We also notice that $\left(H_{\rm K} + 4 \pi M \right)v \sqrt{1-v^{\prime 2}}=\sqrt{H_{\rm K} \left(4 \pi M-2E/M \right)}$. 
Equation (\ref{eq:elliptic_integral_def}) indicates that $w$ is given by 
\begin{equation}
  w
  =
  {\rm sn}
  \left[
    \gamma
    \sqrt{H_{\rm K} \left( 4 \pi M-2E/M \right)}
    t
    +
    \varphi_{0},
    k
  \right],
\end{equation}
where ${\rm sn}(u,k)$ is the Jacobi elliptic function, 
and $\varphi_{0}$ is the initial phase determined by the initial condition. 
Using the relations ${\rm sn}^{2}(u,k)+{\rm cn}^{2}(u,k)=1$ and ${\rm dn}^{2}(u,k)=\sqrt{1-k^{2}{\rm sn}^{2}(u,k)}$, 
we find that the solution of $\mathbf{m}$ on the constant energy curve is given by Eqs. (\ref{eq:mx_sol}), (\ref{eq:my_sol}), and (\ref{eq:mz_sol}). 


The peak frequencies of the Fourier transformation of Eq. (\ref{eq:electric_current_y_total}) in the transverse geometry are discussed as follows. 
We note that Eq. (\ref{eq:electric_current_y_total}) for in-phase synchronization is proportional to $m_{\ell x}(t)m_{\ell y}(t)$  and $m_{\ell z}(t)$. 
Substituting the following formulas \cite{byrd71}, 
\begin{equation}
 {\rm sn}(u,k)
 =
 \frac{2\pi}{k \mathsf{K}(k)}
 \sum_{m=0}^{\infty}
 \frac{q^{m+1/2}}{1-q^{2m+1}}
 \sin
 \left[
   \frac{ \left( 2m+1 \right) \pi u}{2 \mathsf{K}(k)}
 \right],
\end{equation}
\begin{equation}
  {\rm cn}(u,k)
  =
  \frac{2\pi}{k \mathsf{K}(k)}
  \sum_{m=0}^{\infty}
  \frac{q^{m+1/2}}{1+q^{2m+1}}
  \cos
  \left[
    \frac{ \left( 2m+1 \right) \pi u}{2 \mathsf{K}(k)}
  \right],
\end{equation}
\begin{equation}
\begin{split}
  {\rm dn}(u,k)
  =&
  \frac{\pi}{2 \mathsf{K}(k)}
\\
  & +
  \frac{2\pi}{\mathsf{K}(k)}
  \sum_{m=0}^{\infty}
  \frac{q^{m+1}}{1+q^{2(m+1)}}
  \cos
  \left[
    \frac{ \left( m+1 \right) \pi u}{\mathsf{K}(k)}
  \right],
\end{split}
\end{equation}
to Eqs. (\ref{eq:mx_sol}), (\ref{eq:my_sol}), (\ref{eq:mz_sol}), 
where $q=\exp \left[ -\pi \mathsf{K}\left( \sqrt{1-k^{2}} \right)/\mathsf{K}(k) \right]$, 
it is found that the peak frequencies of Eq. (\ref{eq:electric_current_y_total}) appear 
at $f_{n}=(2n-1)f_{0}$ ($n=1,2,3,\cdots$), where the lowest frequency $f_{0}$ is given by Eq. (\ref{eq:frequency}). 


In the longitudinal geometry, Eq. (\ref{eq:electric_current_x_total}) is proportional to 
$m_{1 y}^{2}+m_{2 y}^{2}$. 
When the phase difference of the magnetizations is in-phase or antiphase, 
it becomes $2 m_{1y}^{2}$. 
In this case, using the formula \cite{kiper84}
\begin{equation}
  {\rm dn}^{2}(u,k)
  =
  \frac{\mathsf{E}(k)}{\mathsf{K}(k)}
  +
  \frac{2\pi^{2}}{\mathsf{K}^{2}(k)}
  \sum_{m=1}^{\infty}
  \frac{m q^{m}}{1-q^{2m}}
  \cos
  \left[
    \frac{m \pi u}{\mathsf{K}(k)}
  \right],
\end{equation}
it is found that the Fourier transformation of Eq. (\ref{eq:electric_current_x_total}) has the peaks at $f_{n}=2n f_{0}$. 
On the other hand, when the phase difference is a quarter of a period, 
Eq. (\ref{eq:electric_current_x_total}) is proportional to 
$g(u)\equiv {\rm dn}^{2}(u,k)+{\rm dn} \left[ u+\mathsf{K}(k),k \right] = {\rm dn}^{2}(u,k)+\left[ \left( 1-k^{2} \right)/{\rm dn}^{2}(u,k) \right]$. 
We notice that $g \left[u+\mathsf{K}(k) \right]=g(u)$, indicating that 
the Fourier transformation of Eq. (\ref{eq:electric_current_x_total}) in this case has the peaks at $f_{n}=4n f_{0}$. 


\section{Details of calculations of Eqs. (\ref{eq:Melnikov_T_def}) and (\ref{eq:Melnikov_L_def})}
\label{sec:Details of calculations}

Equations (\ref{eq:Melnikov_T_def}) and (\ref{eq:Melnikov_L_def}) can be calculated 
by substituting the solution of $\mathbf{m}_{\ell}$ on a constant energy curve to the integrals. 
As emphasized in the main text, the phase difference $\Delta\varphi$ between the magnetizations is an important quantity. 
According to Eqs. (\ref{eq:mx_sol}), (\ref{eq:my_sol}), and (\ref{eq:mz_sol}), 
we set $\mathbf{m}_{\ell}$ and $\mathbf{m}_{\ell^{\prime}}$ as  
\begin{equation}
  m_{\ell x}
  =
  \sqrt{
    1
    +
    \frac{2E}{MH_{\rm K}}
  }
  {\rm sn}
  \left[
    \frac{4 \mathsf{K}(k)}{\tau(E)}t,k
  \right],
  \label{eq:mx_sol_1}
\end{equation}
\begin{equation}
  m_{\ell y}
  =
  \sqrt{
    \frac{4\pi M-2E/M}{H_{\rm K}+4\pi M}
  }
  {\rm dn}
  \left[
    \frac{4 \mathsf{K}(k)}{\tau(E)}t,k
  \right],
  \label{eq:my_sol_1}
\end{equation}
\begin{equation}
  m_{\ell z}
  =
  \sqrt{
    \frac{H_{\rm K}+2E/M}{H_{\rm K}+4\pi M}
  }
  {\rm cn}
  \left[
    \frac{4 \mathsf{K}(k)}{\tau(E)}t,k
  \right],
  \label{eq:mz_sol_1}
\end{equation}
and 
\begin{equation}
  m_{\ell^{\prime} x}
  =
  \sqrt{
    1
    +
    \frac{2E}{MH_{\rm K}}
  }
  {\rm sn}
  \left[
    \frac{4 \mathsf{K}(k)}{\tau(E)}t + \Delta\varphi ,k
  \right],
  \label{eq:mx_sol_2}
\end{equation}
\begin{equation}
  m_{\ell^{\prime} y}
  =
  \sqrt{
    \frac{4\pi M-2E/M}{H_{\rm K}+4\pi M}
  }
  {\rm dn}
  \left[
    \frac{4 \mathsf{K}(k)}{\tau(E)}t + \Delta\varphi ,k
  \right],
  \label{eq:my_sol_2}
\end{equation}
\begin{equation}
  m_{\ell^{\prime} z}
  =
  \sqrt{
    \frac{H_{\rm K}+2E/M}{H_{\rm K}+4\pi M}
  }
  {\rm cn}
  \left[
    \frac{4 \mathsf{K}(k)}{\tau(E)}t + \Delta\varphi,k
  \right].
  \label{eq:mz_sol_2}
\end{equation}
The value of $\Delta\varphi$ varies in the rage of $0 \le \Delta\varphi < 4 \mathsf{K}(k)$. 
$\Delta\varphi=0$ corresponds to the in-phase synchronization, 
whereas $\Delta\varphi$ is $2 \mathsf{K}(k)$ for the antiphase synchronization. 


The analytical formulas of Eqs. (\ref{eq:Melnikov_T_def}) and (\ref{eq:Melnikov_L_def}) for the in-phase and antiphase synchronizations 
can be obtained as follows. 
First, since the elliptic functions satisfy 
${\rm sn} \left[ u+2 \mathsf{K}(k),k \right]=-{\rm sn}(u,k)$,
${\rm cn} \left[ u+2 \mathsf{K}(k),k \right]=-{\rm cn}(u,k)$, and 
${\rm dn} \left[ u+2 \mathsf{K}(k),k \right]={\rm dn}(u,k)$,
$\mathscr{W}_{\rm s}^{\rm T}$ has the same magnitude but different sign for $\Delta\varphi=0$ and $\Delta\varphi=2\mathsf{K}(k)$, 
whereas $\mathscr{W}_{\rm s}^{\rm L}$ is the same for the in-phase and antiphase. 
Therefore, it is sufficient to calculate $\mathscr{W}_{\rm s}^{\rm T}$ and $\mathscr{W}_{\rm s}^{\rm L}$ for the in-phase case. 
In this case, it is unnecessary to distinguish $\mathbf{m}_{\ell}$ and $\mathbf{m}_{\ell^{\prime}}$. 
Next, it should be noted that Eq. (\ref{eq:Melnikov_T_def}) includes the following two integrals, 
\begin{equation}
  \oint 
  dt 
  m_{\ell x}^{2}
  m_{\ell y}^{3}
  \propto 
  \int 
  du 
  {\rm sn}^{2}(u,k)
  {\rm dn}^{3}(u,k),
\end{equation}
\begin{equation}
  \oint
  dt 
  m_{\ell x}^{2}
  m_{\ell y}
  m_{\ell z}^{2}
  \propto 
  \int 
  du 
  {\rm sn}^{2}(u,k)
  {\rm cn}^{2}(u,k)
  {\rm dn}(u,k).
\end{equation}
By replacing the integral variable from $u$ with $x={\rm sn}(u,k)$, 
and noting that $d u = dx/\sqrt{ \left( 1-x^{2} \right) \left( 1-k^{2} x^{2} \right)}$, 
these integrals are calculated as 
\begin{equation}
\begin{split}
  &
  \int 
  du 
  {\rm sn}^{2}(u,k)
  {\rm dn}^{3}(u,k)
  =
  \int
  d x 
  \frac{x^{2}(1-k^{2}x^{2})}{\sqrt{1-x^{2}}}
\\
  &=
  \frac{x \sqrt{1-x^{2}} \left[-4 + k^{2} \left( 3+2x^{2} \right) \right] + \left( 4-3 k^{2} \right)\sin^{-1}x}{8},
\end{split}
\end{equation}
\begin{equation}
\begin{split}
  &
  \int
  du
  {\rm sn}^{2}(u,k)
  {\rm cn}^{2}(u,k)
  {\rm dn}(u,k)
  =
  \int 
  dx 
  x^{2}
  \sqrt{1-x^{2}}
\\
  &=
  \frac{x \sqrt{1-x^{2}} \left(-1+2x^{2} \right) + \sin^{-1}x}{8}.
\end{split}
\end{equation}
Using these integrals, Eq. (\ref{eq:Melnikov_T_sol}) is obtained. 
On the other hand, Eq. (\ref{eq:Melnikov_L_def}) includes the following three integrals, 
\begin{equation}
\begin{split}
  \oint 
  dt 
  m_{\ell y}^{3}
  & \propto 
  \int 
  du 
  {\rm dn}^{3}(u,k)
\\
  &=
  \int 
  dx 
  \frac{1-k^{2}x^{2}}{\sqrt{1-x^{2}}}
\\
  &=
  \frac{k^{2} x \sqrt{1-x^{2}} + \left( 2-k^{2} \right) \sin^{-1}x}{2},
\end{split}
\end{equation}
\begin{equation}
\begin{split}
  &
  \oint 
  dt 
  m_{\ell y}^{5}
  \propto 
  \int 
  du 
  {\rm dn}^{5}(u,k)
\\
  &=
  \int 
  dx 
  \frac{\left( 1-k^{2}x^{2} \right)^{2}}{\sqrt{1-x^{2}}}
\\
  &=
  \frac{k^{2} x \sqrt{1-x^{2}} \left[ 8-k^{2} \left( 3+2x^{2} \right) \right]+\left( 8-8k^{2}+3 k^{4} \right)\sin^{-1}x}{8},
\end{split}
\end{equation}
\begin{equation}
\begin{split}
&
  \oint 
  dt 
  m_{\ell y}^{3}
  m_{\ell z}^{2}
  \propto 
  \int 
  du 
  {\rm dn}^{3}(u,k)
  {\rm cn}^{2}(u,k)
\\
  &=
  \int 
  dx 
  \sqrt{1-x^{2}}
  \left(
    1
    -
    k^{2}
    x^{2}
  \right)
\\
  &=
  \frac{x \sqrt{1-x^{2}} \left[4 + k^{2} \left( 1-2x^{2} \right) \right] + \left( 4 - k^{2} \right) \sin^{-1}x}{8}.
\end{split}
\end{equation}
Using these integrals, Eq. (\ref{eq:Melnikov_L_sol}) is obtained. 


When the phase difference is a quarter of a period ($\Delta\varphi=\mathsf{K}(k)$), 
the relations, ${\rm sn} \left[u+\mathsf{K}(k),k \right]={\rm cn}(u,k)/{\rm dn}(u,k)$,
${\rm cn} \left[ u+\mathsf{K}(k),k \right]=-\sqrt{1-k^{2}}{\rm cn}(u,k)/{\rm dn}(u,k)$,
${\rm dn} \left[ u+\mathsf{K}(k),k \right]=\sqrt{1-k^{2}}/{\rm dn}(u,k)$, are used to evaluate Eqs. (\ref{eq:Melnikov_T_def}) and (\ref{eq:Melnikov_L_def}). 
In this case, we notice that all the integrands in Eq. (\ref{eq:Melnikov_T_def}) become odd functions of $t$, 
and the integrals over $0 \le t \le \tau$ are zero. 
Therefore, $\mathscr{W}_{\rm s}^{\rm T}=0$ for $\Delta\varphi=\mathsf{K}(k)$. 
On the other hand, the following integrals are necessary to obtain Eq. (\ref{eq:Melnikov_L_sol_quarter}), 
\begin{equation}
\begin{split}
  \oint 
  dt 
  m_{\ell^{\prime} y}^{2}
  m_{\ell y}
  & \propto
  \int 
  \frac{du}{{\rm dn}(u,k)}
\\
  &=
  \int_{0}^{1}
  \frac{dx}{\sqrt{1-x^{2}} (1-k^{2}x^{2})}
\\
  &=
  \frac{\tan^{-1} \left[ \sqrt{ \left(1-k^{2} \right)/\left( 1-x^{2} \right) }x \right]}{\sqrt{1-k^{2}}},
\end{split}
\end{equation}
\begin{equation}
\begin{split}
  \oint 
  dt 
  m_{\ell^{\prime} y}^{2}
  m_{\ell y}^{3}
  &
  \propto 
  \int 
  du 
  {\rm dn}(u,k)
\\
  &=
  \int 
  \frac{dx}{\sqrt{1-x^{2}}}
\\
  &=
  \sin^{-1}x,
\end{split}
\end{equation}
\begin{equation}
\begin{split}
  &
  \oint 
  dt 
  m_{\ell^{\prime} y}^{2}
  m_{\ell y}
  m_{\ell z}^{2}
  \propto 
  \int 
  du 
  \frac{{\rm cn}^{2}(u,k)}{{\rm dn}(u,k)}
\\
  &=
  \int 
  dx 
  \frac{\sqrt{1-x^{2}}}{1-k^{2}x^{2}}
\\
  &=
  \frac{\sin^{-1}x - \sqrt{1-k^{2}} \tan^{-1} \left[ \sqrt{\left( 1-k^{2} \right)/\left( 1-x^{2} \right) }x \right]}{k^{2}}.
\end{split}
\end{equation}

\end{document}